\documentclass[sigconf]{acmart}

\usepackage{epstopdf}
\usepackage{subcaption}

\usepackage{cleveref}

\crefname{algorithm}{Alg.}{Algs.}
\Crefname{algorithm}{Algorithm}{Algorithms}
\crefname{figure}{Fig.}{Figs.}
\Crefname{figure}{Figure}{Figures}

\usepackage{amssymb,amsfonts,amsmath}

\usepackage[mathscr]{eucal}

\usepackage{stmaryrd}

\usepackage{amsbsy}

\usepackage{bm}

\usepackage{braket}

\usepackage{xparse}

\usepackage{algorithm,algpseudocode}

\algnewcommand\algorithmicparfor{\textbf{parallel for}}
\algdef{S}[FOR]{ParFor}[1]{\algorithmicparfor\ #1\ \algorithmicdo}

\NewDocumentCommand \RangeSet { G{N} } {[\mathinner{#1}]}

\newcommand{\rank}{C}

\NewDocumentCommand \LinearSymbol {} {\oast}
\NewDocumentCommand \LeftSymbol {} {L}
\NewDocumentCommand \RightSymbol {} {R}

\NewDocumentCommand \SingleSize {s O{I} G{n}} {\IfBooleanTF{#1}{\bar}{} {#2}_{#3}}
\NewDocumentCommand \LinearSize {s O{I}} {\IfBooleanTF{#1}{\bar}{} #2^{\LinearSymbol}}
\NewDocumentCommand \LeftSize {s O{I} G{n}} {\IfBooleanTF{#1}{\bar}{} {#2}_{#3}^{\LeftSymbol}}
\NewDocumentCommand \RightSize {s O{I} G{n}} {\IfBooleanTF{#1}{\bar}{} {#2}_{#3}^{\RightSymbol}}
\NewDocumentCommand \UnfoldSize {s O{I} G{n}} {\IfBooleanTF{#1}{\bar}{} {#2}_{\neq #3}}
\NewDocumentCommand \FullSize{s O{I} G{N}} {%
  \IfBooleanTF{#1}%
  {\SingleSize*[#2]{0} \times \cdots \times \SingleSize*[#2]{#3-1}}%
  {\SingleSize[#2]{0} \times \cdots \times \SingleSize[#2]{#3-1}}%
}

\NewDocumentCommand \SizeVec {s O{J}} {\mathbf{\IfBooleanTF{#1}{\bar}{} {#2}}}

\NewDocumentCommand \SingleIndex {s O{j} G{n}} {\IfBooleanTF{#1}{\bar}{} {#2}_{#3}}

\NewDocumentCommand \FullIndex{s O{j} G{N}} {%
  \IfBooleanTF{#1}%
  {(\SingleIndex*[#2]{0}, \SingleIndex*[#2]{1}, \dots, \SingleIndex*[#2]{#3-1})}%
  {(\SingleIndex[#2]{0}, \SingleIndex[#2]{1}, \dots, \SingleIndex[#2]{#3-1})}%
}

\NewDocumentCommand \IndexVec {s O{j}} {\mathbf{\IfBooleanTF{#1}{\bar}{} {#2}}}

\NewDocumentCommand \FullSubscript{s O{j} G{N}} {%
  \IfBooleanTF{#1}%
  {\SingleIndex*[#2]{0} \SingleIndex*[#2]{1} \dots \SingleIndex*[#2]{#3-1}}%
  {\SingleIndex[#2]{0} \SingleIndex[#2]{1} \dots \SingleIndex[#2]{#3-1}}%
}

\NewDocumentCommand \FacMat {s O{U} G{n}} {{\mathbf{\IfBooleanTF{#1}{\bar}{} #2}}^{(#3)}}
\NewDocumentCommand \FacSize {O{I} G{n}} {#1_{#2} \times R_{#2}}

\NewDocumentCommand \Tensor {s O{Y}} {\boldsymbol{\IfBooleanTF{#1}{\bar}{}
{\mathscr{\MakeUppercase{#2}}}}}
\NewDocumentCommand \ColumnBlock {O{\Mz{Y}{n}} G{\ell}} {#1 \left[#2\right]}

\newcommand{\Tra}{{\sf T}}

\newcommand{\V}[2][]{{\bm{#1\mathbf{\MakeLowercase{#2}}}}} 
 
\newcommand{\M}[2][]{{\bm{#1\mathbf{\MakeUppercase{#2}}}}} 
\newcommand{\Mn}[3][]{{\bm{#1\mathbf{\MakeUppercase{#2}}}}_{#3}} 
 
\newcommand{\Mb}[2]{\M{#1}[#2]}

\newcommand{\Mz}[3][]{\M[#1]{#2}_{(#3)}}
\newcommand{\Mzb}[3]{\Mz{#1}{#2}[#3]}

\newcommand{\Khat}{\odot} 
\newcommand{\Hada}{\ast} 
\newcommand{\BigHada}{\mathop{\mbox{\fontsize{18}{19}\selectfont $\circledast$}}} 

\setcopyright{none}

\acmConference[]{}{}{}
\acmYear{2018}
\copyrightyear{2018}

\graphicspath{{fig/}{plots/}{plots/SpUp/}{plots/3alg_seq/}{plots/3alg_par/}{plots/KRP_plts}}

\begin{document}
\title{Shared-Memory Parallelization of MTTKRP for Dense }

\author{Koby Hayashi, Grey Ballard, Yujie Jiang, Michael J. Tobia}
\affiliation{%
  \institution{Wake Forest University}
  \streetaddress{P.O. Box 1234}
  \city{Winston Salem}
  \state{NC}
  \postcode{27109}
}
\email{{hayakb13,ballard,jiany14,tobiamj}@wfu.edu}


\begin{abstract}
The matricized-tensor times Khatri-Rao product (MTTKRP) is the computational bottleneck for algorithms computing CP decompositions of tensors.
In this paper, we develop shared-memory parallel algorithms for MTTKRP involving dense tensors.
The algorithms cast nearly all of the computation as matrix operations in order to use optimized BLAS subroutines, and they avoid reordering tensor entries in memory.
We benchmark sequential and parallel performance of our implementations, demonstrating high sequential performance and efficient parallel scaling.
We use our parallel implementation to compute a CP decomposition of a neuroimaging data set and achieve a speedup of up to $7.4\times$ over existing parallel software.
\end{abstract}

%
%




\maketitle

\section{Introduction}

Tensor decompositions provide a means of data analysis for multi-dimensional data.
In particular, the CP decomposition is a generalization of the matrix singular value decomposition (or principal component analysis), providing a low-rank approximation of data.
This model representation of the data can be used in applications such as blind source separation (interpreting each component as a source signal \cite{CM+15}), in anomaly detection (identifying data points that are not explained by the model \cite{STF06}), and for predicting missing or future data \cite{ADKM11}.
Interest in tensor analysis and the use of the CP decomposition has been growing recently; we refer the reader to survey papers for a more exhaustive list of references \cite{AGHKT14,KB09,SL+17}.

In addition to the growing interest, the increasing size of today's data sets has brought a higher demand for high-performance implementations of the fundamental computational kernels.
For example, nearly all of the time computing CP decompositions occurs in an operation known as matricized-tensor times Khatri-Rao product (MTTKRP).
Most of the available tensor analysis software packages \cite{TensorToolbox,Tensorlab} are written in Matlab, yielding limitations on performance and utility of multicore and other high-performance architectures.
While there have been many recent developments in efficient software for sparse tensor decompositions \cite{LCPSV17,SPK17}, there remain few options in the case of dense tensors, which is the subject of this work.

Our motivating application is a neuroimaging data analysis problem involving functional MRI (fMRI) data.
We are given an input tensor representing correlations between pairs of regions of interest in the brain over time and for various human subjects.
We wish to extract functional brain networks to study how they behave over time relative to a cognitive task and how they relate to and differentiate among subjects.
We discuss this and related problems in more detail in \Cref{sec:application}.
Because of limitations in memory and computational time, the regions of interest are highly coarsened versions of the data.
The existing approach uses the Matlab Tensor Toolbox \cite{TensorToolbox}, but the computational time is a bottleneck in the analysis process.
In order to decrease the time and allow for analysis of larger data sets with finer granularity, our goal is to develop shared-memory parallelizations of the MTTKRP computation in order to utilize multi-core servers.

One advantage of tensor computations is that they can often be cast as matrix operations, which have been well-optimized via the BLAS interface for today's architectures.
In particular, the bulk of MTTKRP corresponds to a single matrix-matrix multiplication.
Unfortunately, using BLAS requires that matrices be stored in regular layouts in memory (e.g., column-major ordering), and it is impossible to choose a dense tensor data layout in memory that is conducive to direct BLAS calls in all cases.
Thus, using BLAS directly requires reordering tensor entries in memory, which is usually too expensive.
The main task in optimizing dense MTTKRP is to employ BLAS in a way that respects a single tensor data layout and avoids tensor reordering.
We discuss MTTKRP in context of the CP decomposition, along with related work, in \Cref{sec:background}.

We consider two MTTKRP algorithms, which we refer to as 1-step and 2-step, that cast the computation as calls to BLAS and never reorder the tensor.
The 1-step algorithm is novel for MTTKRP, using ideas from optimization of a related tensor computation \cite{ABK16,LBPSV15}; the 2-step algorithm was developed by Phan \emph{et al.} \cite{PTC13}.
We also develop a parallel algorithm for computing the Khatri-Rao product of matrices, which is needed for the 1-step and 2-step algorithms.
These sequential and parallel algorithms are presented in \Cref{sec:algorithms}.
We benchmark the algorithms in \Cref{sec:performance}, comparing their performance to baselines, and demonstrating high sequential performance and efficient parallel scaling on a multicore server.

To summarize, the primary contributions of this work are as follows: 
\begin{itemize}
	\item we develop a parallel row-wise algorithm for computing a Khatri-Rao product of multiple matrices;
	\item we implement a new 1-step and an existing 2-step MTTKRP algorithm and parallelize the algorithms using a combination of OpenMP and multithreaded BLAS;
	\item we demonstrate performance improvement over a baseline approach and achieve parallel speedups of up to $12\times$ and $8\times$ over 12 threads; and 
	\item we obtain up to a $7.4\times$ speedup over existing software for computing the CP decomposition of fMRI tensors.
\end{itemize}


\section{Background}
\label{sec:background}

\subsection{Notation}
\label{sec:notation}

Tensors will be denoted using Euler script (e.g., $\Tensor[X]$), matrices will be denoted using upper case bold face type (e.g., $\M{M}$), and vectors will be denoted as lower case bold face type (e.g., $\V{v}$).
We use Matlab-style notation to index into tensors, matrices, and vectors.
For example, $\M{M}(:,c)$ is the $c$th column of matrix $\M{M}$.
Scalar integer values will not be bold-faced, and we use brackets to indicate sets of integers: $\RangeSet{N}=\{0,1,\dots,N{-}1\}$.
Note that we use 0-indexing throughout.
An $N$-dimensional tensor will be referred to as $N$-way or order $N$.
An $N$-way tensor is rank-1 if it can be represented by an outer product of $N$ vectors, one vector in each mode.

We use the notation $\FullSize$ to specify the dimensions of an $N$-way tensor.
For shorthand, we let $I = \prod_{k\in[N]} I_k$ be the total number of entries in the tensor.
We also define $\UnfoldSize = \prod_{n\neq k\in[N]} I_k$ to be the product of all modes but $n$, $\LeftSize = \prod_{n>k\in[N]} I_k$ to be the product of all modes to the left of $n$, and $\RightSize = \prod_{n<k\in[N]} I_k$ to be the product of all modes to the right of $n$.

The \emph{Hadamard} product, or element-wise product, is denoted by $\Hada$.
For example, $\M{C} = \M{A} \Hada \M{B}$ implies $\M{C}(i,j) = \M{A}(i,j) \cdot \M{B}(i,j)$.
The Kronecker product, a generalization of an outer product of vectors, is denoted by $\otimes$.
The \emph{Khatri-Rao} product is denoted by $\Khat$ and will be central to this work.
It can be considered a column-wise Kronecker product, or it can be defined row-wise using the Hadamard product.
Given an $I_A\times \rank$ matrix $\M{A}$ and an $I_B\times\rank$ matrix $\M{B}$, the Khatri-Rao product $\M{K} = \M{A} \odot \M{B}$ has dimension $I_AI_B\times\rank$ (note that the input matrices must have the same number of columns).
Defined column-wise, we have $\M{K}(:,c) = \M{A}(:,c) \otimes \M{B}(:,c)$ for $c\in[C]$.
Defined row-wise, we have $\M{K}(r_B{+}r_AI_B,:) = \M{A}(r_A,:) \Hada \M{B}(r_B,:)$. 

To describe how tensors are stored in memory, we define the standard linearization of tensor entries that generalizes column-major order of matrix entries.
Given a tensor entry $(i_0,\dots,i_{N-1})$, its index in the linearization is given by $\ell = \sum_{n\in[N]} i_n \cdot \LeftSize$.

We also \emph{matricize} or unfold a tensor into a matrix.
A mode-$n$ fiber of a tensor is a vector of entries that share all indices but one; for example, $\Tensor[X](i,:,k)$ is a mode-1 fiber of $\Tensor[X]$.
Arranging all of the mode-$n$ fibers into the columns of a matrix, we obtain the mode-$n$ matricization $\Mz{X}{n}$, which is an $\SingleSize \times \UnfoldSize$ matrix.
The order of the columns corresponds to a linearization of the remaining modes (excluding mode $n$).
We also use a generalization of this concept, assigning multiple modes to the rows of the matrix and the remaining modes to the columns.
In this matricization, an entry's row index corresponds to a linearization of the row modes, and the column index corresponds to a linearization of the column modes.
We use the notation $\Mz{X}{m:n}$ to denote such a matricization with contiguous row modes, where modes $\{m, m{+}1,\dots, n\}$ are the row modes.

Finally, \emph{tensor-times-matrix} (TTM) is denoted by $\times_n$ for mode $n$ and is defined such that $\Tensor[Y] = \Tensor[X] \times_n \M{M}$ is equivalent to $\Mz{Y}{n}=\M{M}^\Tra\Mz{X}{n}$.
When $\M{M}$ is a column vector, we refer to the operation as tensor-times-vector (TTV).

\subsection{CP Decomposition}

\begin{figure}
\includegraphics[width=\columnwidth]{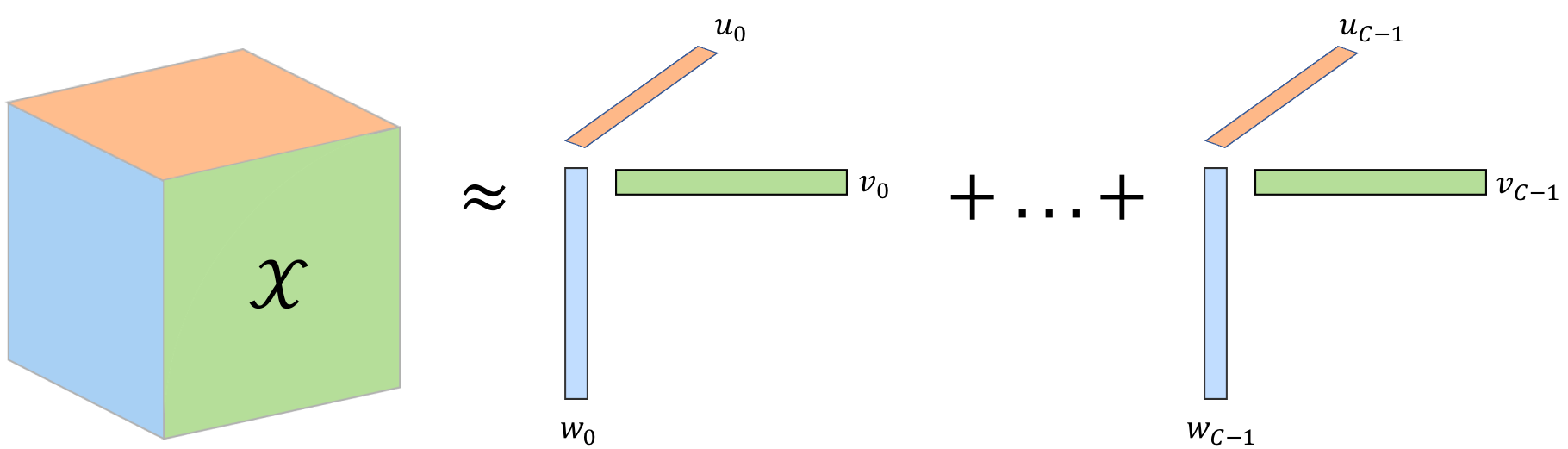}
\caption{Rank-$\rank$ CP decomposition of a 3-way tensor.}
\label{fig:CP}
\end{figure}

A \emph{CP decomposition} is an approximation of a $N$-way tensor $\Tensor[X]$ by a model tensor $\Tensor[Y]$ that is a sum of $\rank$ rank-1 tensors, as shown in \Cref{fig:CP}.
In this section we focus on a 3-way example, but the CP decomposition extends to any $N > 3$ (for more details, see \cite{KB09}, for example).
A CP model is an $N$-way, rank-$\rank$ tensor.
This tensor is represented as a set of $N$ matrices called \emph{factor matrices}.
In general, the $n^{th}$ factor matrix, denoted by $\M{U}_n$, has $I_{n}$ rows and $\rank$ columns.
In our 3-way example, the model has factor matrices $\M{U}, \M{V},$ and $\M{W}$, and its entries are given by
$$\Tensor[Y](i,j,k) = \sum_{c\in[\rank]} \M{U}(i,c) \cdot \M{V}(j,c) \cdot \M{W}(k,c).$$
We also use the notation $\Tensor[Y]=\llbracket \M{U}, \M{V}, \M{W} \rrbracket$.

One commonly used method of computing the CP decomposition is the Alternating Least Squares (CP-ALS) method.
In CP-ALS, one factor matrix is updated at a time and the rest are kept fixed.
This update involves solving a linear least squares problem that can be expressed in matrix notation; for example, the update of $\M{V}$ has the form $\M{v} =  \Mz{X}{1} (\M{u} \Khat \M{W} )(\M{u}^\Tra \M{u} \Hada \M{W}^\Tra \M{W})^\dagger$.

In general, each factor matrix update consists of 3 operations:
\begin{itemize}
	\item Matricized Tensor Times Khatri Rao Product (MTTKRP),
	$$ \M{M} = \Mz{X}{n}(\M{U}_{N-1} \odot \cdots \odot \M{U}_{n+1} \odot  \M{U}_{n-1} \odot \cdots \odot \M{U}_{0}),$$
	\item Gram matrix and Hadamard product computation,
	$$\M{H} = \BigHada_{n\neq k \in [N]} (\M{U}_k)^\Tra(\M{U}_k)$$
	\item and solving a linear system: $\M{U}_{n} = \M{M} \M{H}^\dagger, $
\end{itemize}
where $\dagger$ denotes the pseudoinverse.
Ignoring the cost of forming the Khatri-Rao product matrix (KRP), the number of flops required to multiply $\Mz{X}{n}$ by the KRP is $O(I\rank)$.
The number of flops required to compute $\M{H}$ is $O(\rank^2N+\rank\sum_{n\neq k \in[N]} I_k^2)$, and the number of flops involved in solving the linear system is $O(\rank^3+I_n\rank^2)$.
Thus, as $I$ is the product of all the tensor dimensions, in the typical case nearly all of the computation is spent in MTTKRP.
We note that there are alternative optimization schemes to CP-ALS, but because MTTKRP is part of the gradient, nearly all of them require computing and are bottlenecked by MTTKRP.

\subsection{MTTKRP}

The most straightforward way to compute the MTTKRP is to form the matricized tensor explicitly (as a column- or row-major matrix), form the KRP explicitly, and then use a BLAS call to perform the matrix multiplication \cite{BK07}.
This approach benefits from an efficient matrix multiplication.
However, forming an explicit matricized tensor involves reordering the tensor entries (for most modes), which is a completely memory-bound operation and can become the bottleneck.
Likewise, the KRP computation involves $O(\UnfoldSize\rank)$ flops to produce a matrix of size $\UnfoldSize\times\rank$, which has a very low arithmetic intensity and will also be memory bound.
The goal in this paper is to avoid reordering tensor entries and perform the KRP computation and matrix multiplication as fast as possible.

\subsection{Related Work}

There are several Matlab software packages that implement optimization techniques for computing CP decompositions and include functions for computing MTTKRP, including N-way Toolbox \cite{AB00}, Tensor Toolbox \cite{TensorToolbox}, and Tensorlab \cite{Tensorlab}.
Recently, there have been efforts to develop more efficient implementations of MTTKRP to compute CP decompositions of sparse tensors, involving parallelizations for multi-core, many-core, and distributed-memory systems.
Kaya and U\c{c}ar \cite{KU15} develop a distributed-memory implementation of CP (and MTTKRP) for sparse tensors using hypergraph partitioning techniques to optimize performance.
Smith \emph{et al.} \cite{SRSK15} develop SPLATT, a shared-memory parallel library for sparse CP, which has been extended to distributed-memory \cite{SK16} and many-core platforms \cite{SPK17}.
Li \emph{et al.} \cite{LCPSV17} present AdaTM, a shared-memory parallel framework for sparse CP that reuses intermediate quantities across the MTTKRPs in different modes to save computation.

In the dense case, Bader and Kolda \cite{BK07} proposed the straightforward approach described in the previous section.
Phan \emph{et al.} \cite{PTC13} introduce an alternative approach that avoids reordering tensor entries but still casts most of the computation in terms of matrix multiplication.
We implement sequential and parallel versions of their algorithm in \Cref{sec:algorithms}.
They also show how to avoid redundant computation across MTTKRPs for different modes, but we do not consider that optimization in this paper.
Vannieuwenhoven \emph{et al.} \cite{VMV15} also implement the algorithm of Phan \emph{et al.} and combine it with a blocking approach to minimize the temporary memory footprint of a dense MTTKRP.
They show that minimizing the memory footprint does not have an adverse effect on performance, though they do not parallelize the algorithm.

Other related work exploits the data layout of matricized tensors and avoid reordering tensor entries using similar ideas to ours for a different tensor computation, known as tensor-times-matrix (TTM).
Li \emph{et al.} \cite{LBPSV15} develop a parallelization framework for computing TTMs with dense tensors on multicore platforms.
Austin \emph{et al.} \cite{ABK16} present distributed-memory parallel algorithms for computing the Tucker decomposition of dense tensors, which includes the sequential implementation of TTM that avoids reordering entries.

Other parallelizations of the CP decomposition and MTTKRP for dense tensors include those of Liavas \emph{et al.} \cite{LK+17} and Aggour and Yenner \cite{AY16}.
Liavas \emph{et al.} consider the nonnegative case and implement a distributed-memory parallel MTTKRP, presenting performance results for 3-way tensors of equal-sized dimensions.
Aggour and Yenner also use distributed-memory parallelization and focus on 3-way tensors that have a single long dimension.
We are unaware of any work that parallelizes MTTKRP for dense tensors on shared-memory platforms.

\section{Neuroimaging Application}
\label{sec:application}

Our motivation for this work is a need for more efficient software to extract brain connectivity information from functional Magnetic Resonance Imaging (fMRI) data.
The CP decomposition is a useful tool for neuroimaging research in general because it affords a multidimensional approach to the analysis of large and potentially heterogeneous data sets.  
It allows researchers to extract commonalities from diverse groups of human data and generate dynamic brain connectivity maps \cite{HW+13}.

In our case, for example, subjects are given a cognitive task that lasts several minutes, and fMRI information is gathered for a discrete set of voxels in the brain.
The voxel information is aggregated into regions of interest, and then for each subject and at each time point, the instantaneous correlations between all pairs of regions of interest are computed.
This produces a time-by-subject-by-region-by-region dense tensor, and we use a CP decomposition to extract components that represent brain networks that vary over time and have varying representation over subjects.
Analysis of these components yields a more complete picture of how brain connectivity relates to tasks and individual performance, and it can be used to differentiate among individuals. 
Such advanced data analyses hold promise to reveal important early onset symptoms of neurogenerative disease so that prophylactic treatments may be devised.

The spatial and temporal resolution of human functional neuroimaging data is always improving due to developments in MRI technology.  
MRI techniques that push the limits of achievable spatial and temporal resolution lead to larger and richer brain imaging data sets, which will rely on efficient algorithms for analysis.  
In addition to the ever-improving spatial and temporal resolution of human functional MRI data collection, research studies are employing larger and larger group sizes as well---referred to as population imaging studies---with the aim of building a data driven discovery science of the human mind and brain (e.g., Human Connectome Project \cite{VEU+12}, UK BioBank \cite{MA-A+16}).  
The increasing sample sizes in combination with the improving MRI data collection requires efficient and scalable methods for analysis. 
The need to discover the optimal rank of multi-way and multi-modal data, and employ multiple random starts to ensure uniqueness, reliability, and reproducibility of the solutions for the massive data sets, all implicate large-scale computing as crucial to the success and advancement of these and similar projects.


\section{Algorithms}
\label{sec:algorithms}


\subsection{Khatri-Rao Product}

We consider the Khatri-Rao product (KRP) of $Z$ matrices.
In the context of MTTKRP, the output for the $n$th mode mathematically depends on the \emph{full KRP}:
$$\M{K} = \Mn{U}{N-1} {\Khat} {\cdots} {\Khat} \Mn{U}{n+1} \Khat \Mn{U}{n-1} {\Khat} {\cdots} {\Khat} \Mn{U}{0}.$$
However, we consider MTTKRP algorithms that form the full KRP as well as those that do not form it explicitly and instead compute partial KRPs, such as the \emph{left KRP}:
$$\M{K}_L =  \M{U}_{(0)} \Khat ... \Khat \M{U}_{(n-1)},$$
and the \emph{right KRP}:
$$\M{K}_R = \M{U}_{(n+1)} \Khat ... \Khat \M{U}_{(N-1)}.$$
Here, we consider the generic case of $Z$ input matrices.

The KRP is often defined as column-wise Kronecker product, and it can be computed that way.
However, we consider a row-wise approach that is more easily parallelized.
Recall that each row of a KRP is the Hadamard product of a set of rows, one for each input factor matrix.
For example, the $j$th row of $\M{K}=\M{A}\Khat \M{B} \Khat \M{C}$ can be expressed as $\M{K}(j,:)=\M{A}(a,:) \Hada \M{B}(b,:) \Hada \M{C}(c,:)$, where $j = aI_BI_C+bI_C+c$ and $I_B$ and $I_C$ are the number of rows of $\M{B}$ and $\M{C}$, respectively.

Naively, a KRP of $Z$ matrices requires $Z{-}1$ Hadamard products per row of the output matrix.
This number can be reduced by storing and re-using partially computed Hadamard products.
In the example above, $\M{A}(a,:)\Hada \M{B}(b,:)$ will be used $I_{C}$ times in computing $\M{K}$.
Saving this partial Hadamard product when it is first computed allows for reuse in computing subsequent rows of \M{K}.
For a KRP of $Z{\geq} 3$ matrices, one can store $Z{-}2$ Hadamard products to save computation and perform roughly one Hadamard product per row of the output matrix.

\subsubsection{Sequential}

\Cref{alg:KRP} presents the pseudocode that implements this technique, computing the output matrix $\M{K}$ one row at a time and re-using partial Hadamard products.
The vector $\V{\ell}$ is a multi-index that stores the row indices of the input matrices that correspond to a row of the output.
The matrix $\M{P}$ is a temporary matrix that stores the $Z-2$ intermediate Hadamard products, which are each of length $\rank$, the number of columns of the input matrices.
For the sequential algorithm, $\ell$ is initialized to $\V{0}$ in line 2, $\M{P}(0,:)$ is initialized to $\M{U}_0(0,:)\Hada \M{U}_1(0,:)$ and $\M{P}(z,:)$ is initialized to $\M{P}(z{-}1,:)\Hada \M{U}_{z{+}1}(0,:)$ for $0<z\leq Z{-}3$ in line 3.
Inside the for loop, the multi-index is incremented at line 6.
For every index that changes (except for the last one), the corresponding temporary Hadamard products in $\M{P}$ must be re-computed (at line 7).
However, this update occurs infrequently -- extra computation is required only one out of every $I_{Z{-}1}$ iterations.
Thus, the dominant cost is that of the single Hadamard product of line 5, which occurs once per row.

\subsubsection{Parallel}

Parallelizing \cref{alg:KRP} is straightforward, so we do not provide pseudocode.
The parallel variant works as follows.
We assign the rows of the output matrix to threads in contiguous blocks.
Each thread initializes $\ell$ and $\M{P}$ according to its starting row index (rather than starting with row 0).
Then the algorithm proceeds exactly as in the sequential case, except that the thread stops iterating after it computes its last assigned row.

\begin{algorithm}[t]
\caption{Row-wise Khatri-Rao Product with Reuse}
\label{alg:KRP}
\begin{algorithmic}[1]
\Require  $\M{U}_z$ is an $J_z \times \rank$ matrix, for $z\in \RangeSet{Z}$, $Z\geq 3$
\Procedure{$\M{K} = $ KRP}{$\M{U}_0,\dots,\M{U}_{Z-1}$}
	\State initialize$(\V{\ell})$ \Comment{Initialize multi-index of length $Z$} \label{line:KRP:initindex}
	\State initialize$(\M{P})$	\Comment{$Z{-}2\times \rank$ matrix for intermediate products} \label{line:KRP:initHP}
	\For{$j \in \RangeSet{\prod J_z}$ }	 \Comment{Loop over rows of output $\M{K}$}
		\State $\M{K}(j,:) \gets \M{P}(Z{-}3,:) \Hada \M{U}_{Z-1}(\ell_{Z-1},:)$
		\State increment$(\V{\ell})$
		\State update$(\M{P})$ \Comment{Update intermediate products if needed}
	\EndFor
\EndProcedure
\Ensure $\M{K}  = \M{U}_{0} \Khat \cdots \Khat \M{U}_{Z-1}$ is $\prod J_z \times \rank$ matrix
\end{algorithmic}
\end{algorithm}

\subsection{1-Step MTTKRP}

We now consider a 1-step approach to compute the MTTKRP.
Given a matricized tensor and an explicit KRP, the idea is to perform the matrix multiplication efficiently.
The benefit of this approach is that most of the computation is cast as matrix multiplication, for which high-performance implementations exist (via the BLAS interface).
However, the principal complication is that the matricizations for all internal modes ($0<n<N{-}1$) of a tensor whose elements are linearized in a natural way are not column- or row-major in memory, which is a requirement for the BLAS interface.
The time required to re-order tensor elements to obtain a column- or row-major matricization is usually prohibitive, negating the benefit of BLAS performance.

\begin{figure}
\includegraphics[width=\columnwidth]{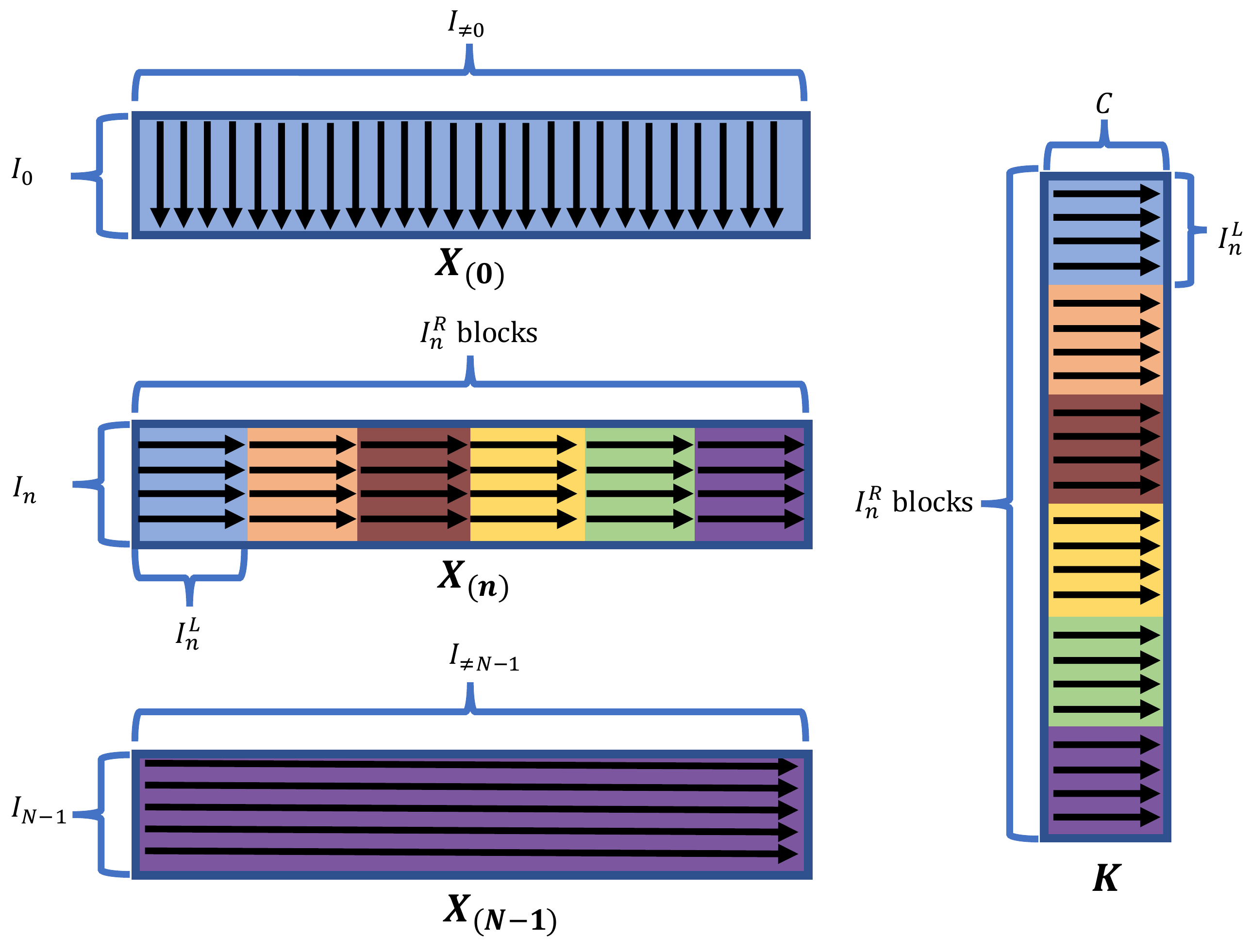}
\caption{Data layout of the matricizations of an $N$-way tensor $\Tensor[X]$.
Note that  $\Mz{X}{0}$ is column-major, $\Mz{X}{N-1}$ is row-major, and $\Mz{X}{n}$ consists of $\RightSize$ contiguous row-major submatrices of dimension $\SingleSize\times\LeftSize$.
A conformal partitioning of the $n$th-mode KRP for the $n$th mode MTTKRP is depicted on the right.}
\label{fig:1step}
\end{figure}

Our main idea of 1-Step MTTKRP is to perform the matrix multiplication without reordering tensor entries, using multiple BLAS calls.
Our algorithm is based on the observation that given the natural linearization of tensor entries, the $n$th mode matricization can be seen as a contiguous set of submatrices, each of which is stored row-major in memory \cite{LBPSV15,ABK16}.
\Cref{fig:1step} shows how $\Mz{X}{n}$ is ordered in memory, and it also shows how the KRP matrix $\M{K}$ can be conformally partitioned to perform the matrix multiplication as a block inner product.

\subsubsection{Sequential}

\Cref{alg:seq-1s-MTTKRP} shows the pseudocode for the sequential 1-step algorithm.
The first step is to compute the full KRP using \Cref{alg:KRP}.
In the case of mode 0, $\Mz{X}{0}$ is column-major so MTTKRP can be performed with a single BLAS call (line 4).
For other modes, to avoid reordering tensor entries, the algorithm may have to make multiple BLAS calls.
As shown in \cref{fig:1step}, lines 6 and 7 define a conformal partitioning of the two matrices so that the MTTKRP can be performed as a block inner product (the sum of submatrix multiplies).
This partitioning is such that each submatrix of the matricization (of size $\SingleSize\times\LeftSize$) and each submatrix of the KRP (of size $\LeftSize\times\rank$) is row-major in memory.
Thus, each multiplication in line 9 is a BLAS call.

\begin{algorithm}
\caption{Sequential 1-step MTTKRP}
\label{alg:seq-1s-MTTKRP}
\begin{algorithmic}[1]
\Require $\Tensor[X]$ is $\FullSize[I]$, $\Tensor[Y]= \llbracket \M{U}_{0}, \dots, \M{U}_{N-1} \rrbracket$, $n\in \RangeSet{N}$
\Procedure{$\M{M} = $ MTTKRP-1Step-Seq}{$\Tensor[X]$, $\Tensor[Y]$, $n$}
	\State $\M{K} \gets \text{KRP}(\Mn{U}{N-1}, \dots, \Mn{U}{n+1}, \Mn{U}{n-1}, \dots, \Mn{U}{0})$ \Comment{\cref{alg:KRP}}
	\If{$n = 0$} 
		\State $\M{M} \leftarrow \Mz{X}{0} \cdot \M{K}$ \Comment{$\Mz{X}{0}$ is column-major}
	\Else
		\State Partition $\Mz{X}{n}$ into $\RightSize$ column blocks of size $I_n\times \LeftSize$
		\State Partition $\M{K}$ into $\RightSize$ row blocks of size $\LeftSize \times \rank$
		\For{$j \in \RangeSet{\RightSize[I]}$} \Comment{Loop over column blocks of $\Mz{X}{n}$}
		\State $\M{M} \gets \M{M} + \Mzb{X}{n}{j} \cdot \Mb{K}{j}$ \Comment{$\Mzb{X}{n}{j}$ is row-major}
	\EndFor
	\EndIf
\EndProcedure
\Ensure $\M[M]\ =\ \Mz{X}{n} \cdot (\Mn{U}{N-1} {\Khat} {\cdots} {\Khat} \Mn{U}{n+1} \Khat \Mn{U}{n-1} {\Khat} {\cdots} {\Khat} \Mn{U}{0})$ is $I_n \times \rank$
\end{algorithmic}
\end{algorithm}

\subsubsection{Parallel}

We use two different techniques to parallelize 1-step MTTKRP, depending on the mode.
We distinguish between external ($n=0$ or $n=N{-}1$) and internal ($0<n<N{-}1$) modes.

For external modes, we parallelize over the \emph{columns} of the matricization.
Each thread is assigned a contiguous set of columns of $\Mz{X}{n}$.
In line 7, the thread computes the corresponding rows of the KRP $\M{K}$ (using a variant of \Cref{alg:KRP}), and in line 8 it performs the multiplication with a BLAS call.
Each thread computes a contribution to the output matrix, so a parallel reduction is performed at the end of the algorithm (line 23).

For internal modes, we parallelize over the \emph{blocks} of the matricization.
In this case, each thread is assigned a set of $\SingleSize\times\LeftSize$ blocks.
The row block of $\M{K}$ that corresponds to the $j$th column block of $\Mz{X}{n}$ is the Khatri-Rao product of the left KRP matrix and the $j$th row of the right KRP matrix.
Thus, $\M{K}_L$ is pre-computed in parallel in line 11, and each thread computes the corresponding row of $\M{K}_R$ (line 14) and Khatri-Rao product (line 15) to obtain the necessary row blocks of $\M{K}$.
The matrix multiplication of line 16 is performed with a BLAS call because each block is row-major.
Again, a parallel reduction is necessary at the end of the algorithm.

Note that the internal-mode parallelization scheme assumes that the number of threads is much less than $\RightSize$ in order to achieve load balance.
We expect this to hold in nearly all cases because $\RightSize$ is a product of tensor dimensions.
If this is not the case (say $\RightSize$ corresponds to only one mode of very small dimension), then an alternative approach would be to use the sequential algorithm with multithreaded BLAS.

\begin{algorithm}
\caption{Parallel 1-Step MTTKRP}
\label{alg:par-1s-MTTKRP}
\begin{algorithmic}[1]
\Require $\Tensor[X]$ is $\FullSize[I]$, $\Tensor[Y]= \llbracket \M{U}_{0}, \dots, \M{U}_{N-1} \rrbracket$, $n\in \RangeSet{N}$
\Require $T$ is the number of threads
\Procedure{$\M{M} = $ MTTKRP-1Step-Par}{$\Tensor[T]$, $\Tensor[Y]$, $n$, $T$}
	\If{$n = 0$ or $n = N-1$}
	\State $b \gets \lceil \UnfoldSize / T \rceil$
	\State Partition $\Mz{X}{n}$ into $T$ column blocks of size $I_n\times b$
	\State Partition KRP $\M{K}$ into $T$ row blocks of size $b\times \rank$
	\ParFor{$t \in \RangeSet{ T }$, private($\overline{\M{M}}_t, \overline{\M{K}}[t]$)}
	\State Compute $\overline{\M{K}}[t]$ \Comment{Variant of \cref{alg:KRP}}
	\State $\overline{\M{M}}_t \leftarrow \M{X}_{(n)}[t] \cdot \overline{\M{K}}[t]$
	\Statex \Comment{$\M{X}_{(n)}[t]$ is submatrix of column- or row-major matrix}
	\EndFor
	\Else
	\State $\M{K}_L \gets \text{KRP}(\Mn{U}{n-1}, \dots, \Mn{U}{0})$ \Comment{Parallel variant of \cref{alg:KRP}}
	\State Partition $\Mz{X}{n}$ into $\RightSize$ column blocks of size $I_n\times \LeftSize$
	\ParFor{$j \in \RangeSet{\RightSize}$, private($\overline{\M{M}}_t,\overline{\M{K}}_t$)} 
	\State Compute $\M{K}_R(j,:)$ \Comment{$j$th row of $\M{K}_R$}
	\State $\overline{\M{K}}_t \leftarrow \M{K_R}(j,:) \Khat \M{K}_L$
	\State $\overline{\M{M}}_t \leftarrow \overline{\M{M}}_t + \M{X}_{n}[j] \cdot \overline{\M{K}}_t$ \Comment{$\M{X}_{n}[j]$ is row-major}
	\EndFor
	\EndIf
	\State $\M{M} \gets \sum_t \overline{\M{M}}_t$ \Comment{Parallel reduction}
\EndProcedure
\Ensure $\M{M}\ =\ \Mz{X}{n} \cdot (\Mn{U}{N-1} {\Khat} {\cdots} {\Khat} \Mn{U}{n+1} \Khat \Mn{U}{n-1} {\Khat} {\cdots} {\Khat} \Mn{U}{0})$ is $\SingleSize \times \rank$
\end{algorithmic}
\end{algorithm}

\subsection{2-Step MTTKRP}

\begin{figure}
	\begin{subfigure}{\columnwidth}
	\includegraphics[width=\columnwidth]{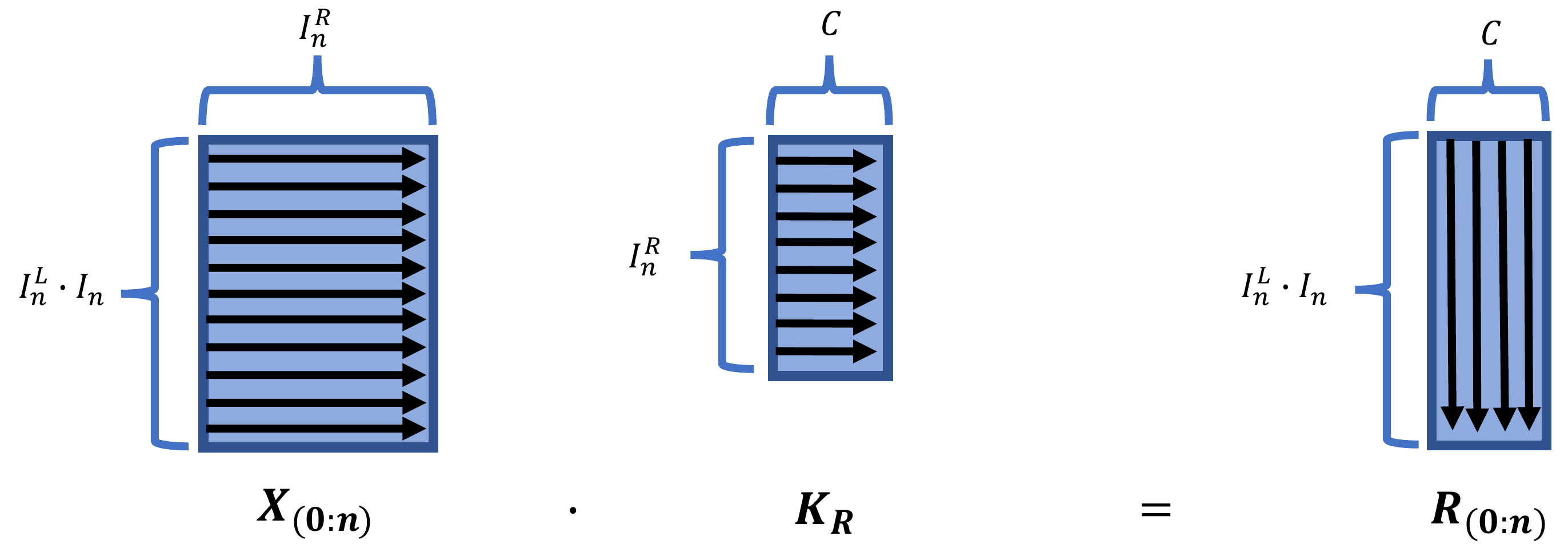}
	\caption{Right partial MTTKRP involving $\Tensor[X]$ and right KRP $\M{K}_R$}
	\label{fig:2step:R1}
	\end{subfigure}
	\begin{subfigure}{\columnwidth}
	\includegraphics[width=\columnwidth]{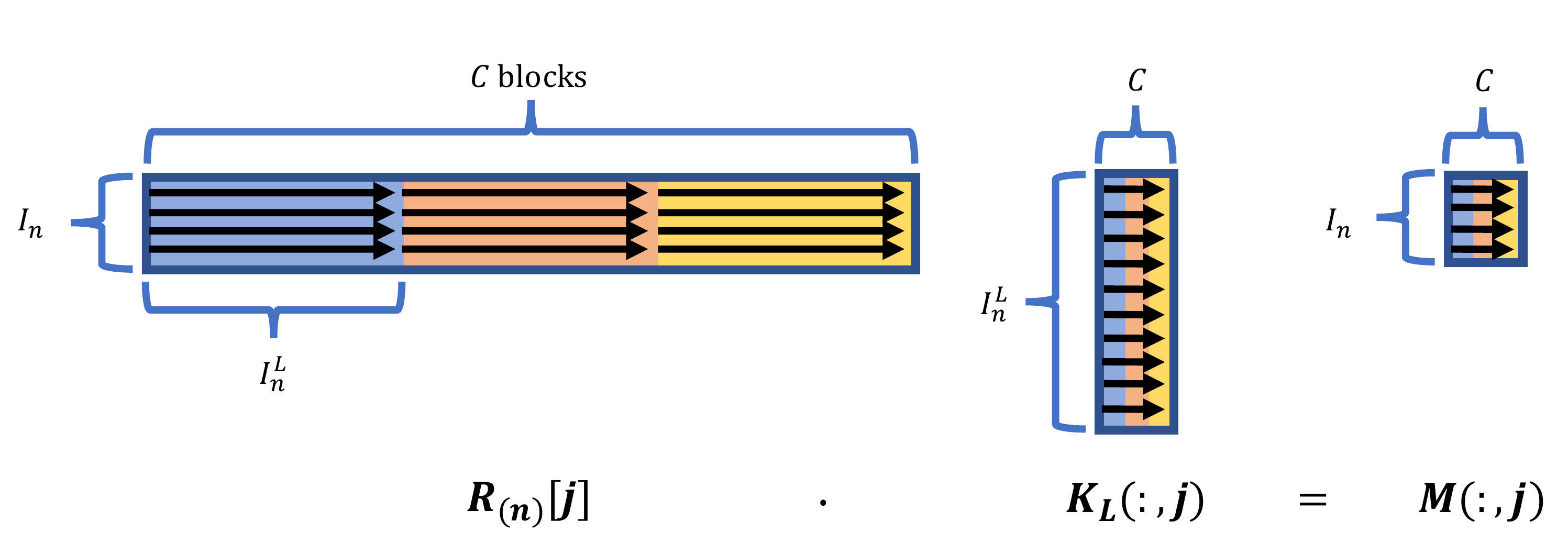}
	\caption{Multi-TTV involving $\Tensor[R]$ and left KRP $\M{K}_L$}
	\label{fig:2step:R2}
	\end{subfigure}
	\begin{subfigure}{\columnwidth}
	\includegraphics[width=\columnwidth]{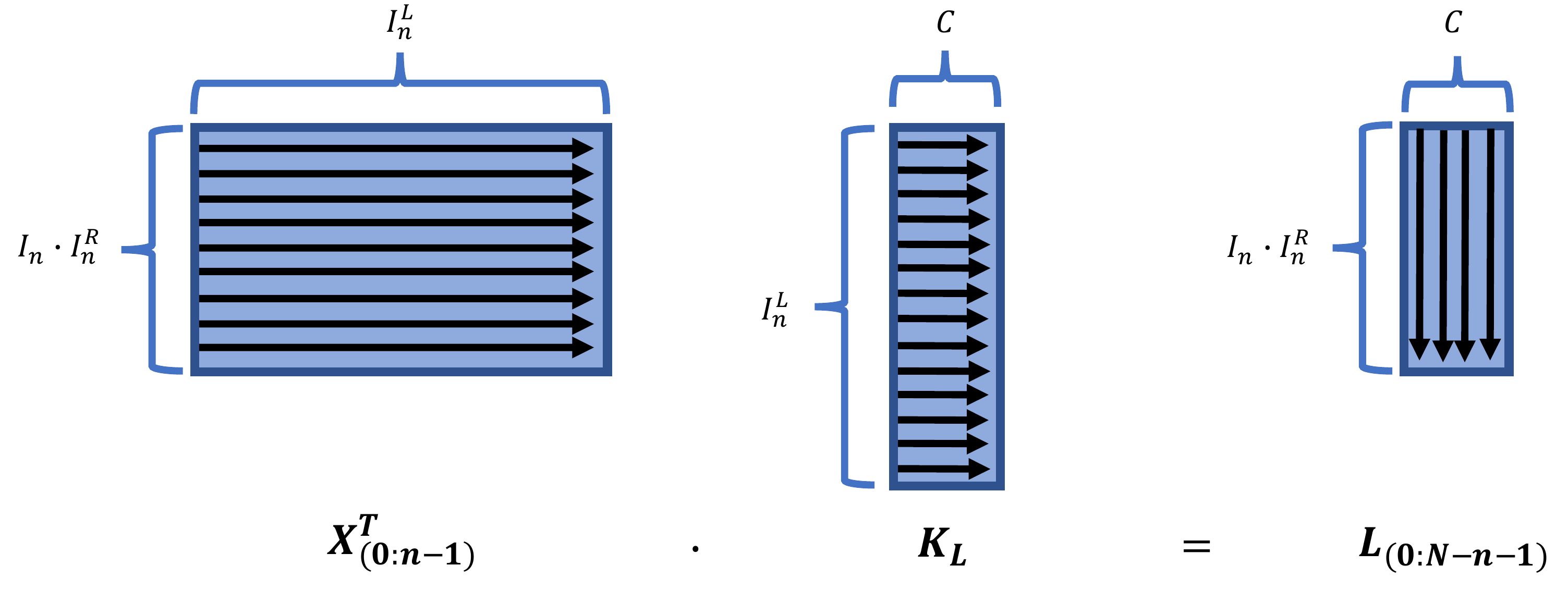}
	\caption{Left partial MTTKRP involving $\Tensor[X]$ and left KRP $\M{K}_L$}
	\label{fig:2step:L1}
	\end{subfigure}
	\begin{subfigure}{\columnwidth}
	\includegraphics[width=\columnwidth]{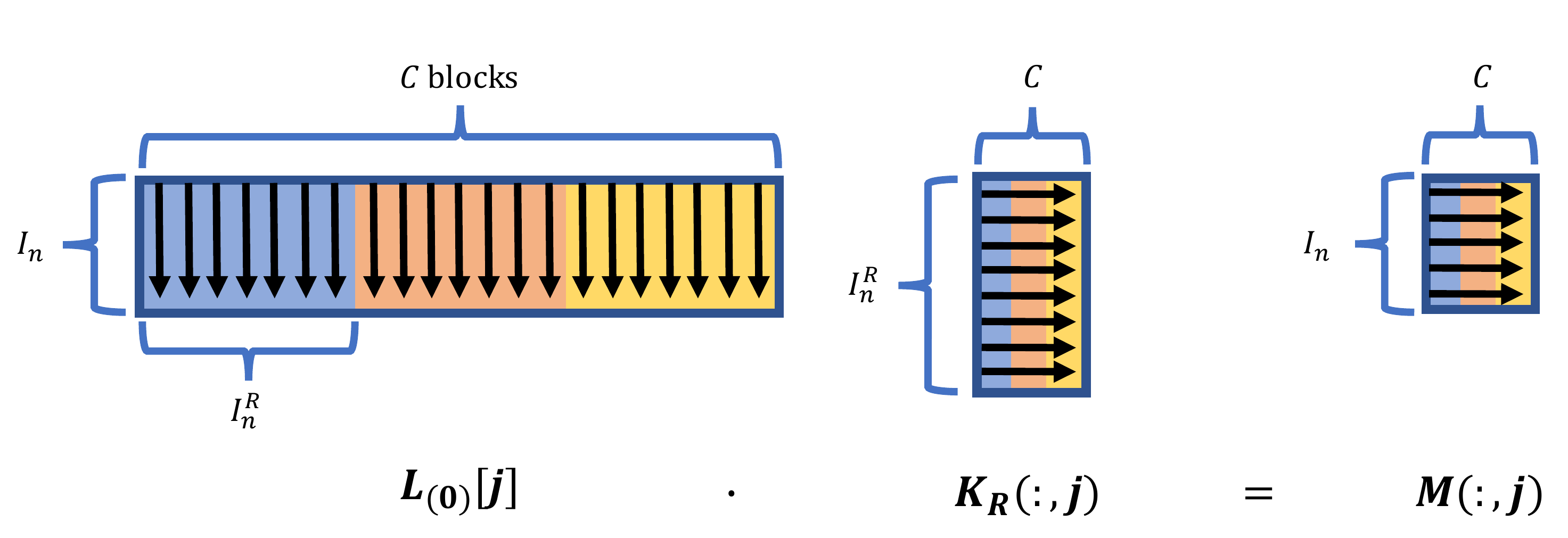}
	\caption{Multi-TTV involving $\Tensor[L]$ and right KRP $\M{K}_R$}
	\label{fig:2step:L2}
	\end{subfigure}
\caption{Data layouts of tensors and matrices involved in computations of 2-step MTTKRP.
\cref{fig:2step:R1,fig:2step:R2} correspond to the 2 steps with right-left ordering; \cref{fig:2step:L1,fig:2step:L2} correspond to the 2 steps with left-right ordering. }
\label{fig:2step}
\end{figure}

The 2-step algorithm first performs a \emph{partial MTTKRP} and then finishes the computation with multiple tensor-times-vector operations (\emph{multi-TTV}).
The algorithm was developed by Phan \emph{et al.} \cite[Section III.B]{PTC13}, but we explain it again here using our notation.
For external modes, the 2-step algorithm degenerates to the 1-step algorithm.
The pseudocode appears in \Cref{alg:2s-MTTKRP}, and the data layouts of each computation are show in \Cref{fig:2step}.

As in the case of the 1-step algorithm, our goal will be to perform the computation without reordering tensor entries.
The first observation is that more general matricizations of the tensor are column-major in memory.
That is, using the notation defined in \Cref{sec:notation}, $\Mz{X}{0:n}$ is column major in memory for all $n$.
This implies that we can compute the matrix product of $\Mz{X}{0:n}$ and $\M{K}_R$ (the right KRP) with a single BLAS call, as shown in \Cref{fig:2step:R1}.
Because this matrix multiplication involves all the tensor entries but only a subset of the input matrices, we refer to this operation as a partial MTTKRP.
(Note that when $n=0$, it is a full MTTKRP.)

The output of a partial MTTKRP is an intermediate quantity which must be combined with the remaining input matrices to obtain the final MTTKRP output.
We interpret the output of this partial MTTKRP as a tensor of dimension $n+2$, defining
$$\Mz{R}{0:n} = \Mz{X}{0:n} \cdot \M{K}_R,$$
so that $\Tensor[R]$ has dimensions $I_0 \times \cdots \times I_n \times \rank$.
Given $\Tensor[R]$, the $j$th column of the MTTKRP output matrix $\M{M}$ is a tensor-times-vector (TTV) operation involving the $j$th subtensor of $\Tensor[R]$ and the $j$th columns of the remaining input matrices:
$$\M{M}(:,j) = \Tensor[R](:,\dots,:,j) \times_0 \M{U}_0(:,j) \cdots \times_{n-1} \M{U}_{n-1}(:,j).$$
Because the operation must be performed for each column of $\M{M}$, we refer to the overall 2nd step as a multi-TTV.

We note that the operation can also be interpreted as an MTTKRP involving subtensor $\Tensor[R](:,\dots,:,j)$ and the set of columns.
Because the subtensor involves a leading set of modes, it is stored as a tensor in natural order, and we can apply the same computational techniques.
In particular, this MTTKRP is done with respect to the last mode of the subtensor, so it requires only one BLAS call.
However, the KRP has only one column in this case, so the BLAS call is for matrix-vector rather than matrix-matrix multiplication.
The matrix-vector product must be performed for each of the $\rank$ output columns, as shown in \Cref{fig:2step:R2}.

The 2 steps described above incorporate the modes to the right in the 1st step and the modes to the left in the 2nd step, but that order can be reversed.
To compute the partial MTTKRP involving the left modes, we observe that $\Mz{X}{0:n-1}^\Tra$ is row major in memory, and we can compute a different temporary quantity
$$\Mz{L}{0:N-n-1} = \Mz{X}{0:n-1}^\Tra\cdot \M{K}_L,$$
where $\Tensor[L]$ is $\SingleSize\times\cdots\times I_{N-1}\times \rank$ (see \Cref{fig:2step:L1}).
The second step multi-TTV involves subtensors of $\Tensor[L]$ and columns of $\M{K}_R$.
The $j$th column of the output is given by
$$\M{M}(:,j) = \Tensor[L](:,\dots,:,j) \times_{n+1} \M{U}_{n+1}(:,j) \cdots \times_{N-1} \M{U}_{N-1}(:,j).$$
Computationally, we can interpret each TTV as an MTTKRP with respect to the first mode of the subtensor, so again it involves only one BLAS call for each matrix-vector multiplication (see \Cref{fig:2step:L2}).

The pseudocode for the 2-step algorithm is given in \Cref{alg:2s-MTTKRP}.
It starts by computing both left and right partial KRPs.
Given that either ordering of the steps is correct, the algorithm chooses the ordering that minimizes computation in the 2nd step (the number of flops in the 1st step is the same).
If it is more efficient to do the left side first, it computes the left partial MTTRKP in line 5 and the multi-TTV in lines 6--9.
Otherwise, it computes the right partial MTTKRP in line 11 and the multi-TTV in lines 12--15.

The bulk of the computation occurs in the partial MTTKRP, which is a matrix multiplication requiring a single BLAS call.
We note that the dimensions of this matrix multiplication are more balanced than in the case of the full MTTKRP.
Parallelization of this algorithm is all within the BLAS calls, so \Cref{alg:2s-MTTKRP} applies for both sequential and parallel variants.

\begin{algorithm}[t]
  \caption{Sequential/Parallel 2-step MTTKRP}
  \label{alg:2s-MTTKRP}
  \begin{algorithmic}[1]
  \Require $\Tensor[X]$ is $\FullSize[I]$, $\Tensor[Y]= \llbracket \M{U}_{0}, \dots, \M{U}_{N-1} \rrbracket$, $n\in \RangeSet{N}$
  \Procedure{$\M{M} = $ MTTKRP-2step}{$\Tensor[X]$, $\Tensor[Y]$, $n$}
  	\State $\M{K}_L \gets \text{KRP}(\Mn{U}{n-1}, \dots, \Mn{U}{0})$ \Comment{\cref{alg:KRP}, Left Partial KRP}
	\State $\M{K}_R \gets \text{KRP}(\Mn{U}{N-1}, \dots, \Mn{U}{n+1})$ \Comment{\cref{alg:KRP}, Right Partial KRP}
	\If{$\LeftSize > \RightSize$} \Comment{Use Left Partial MTTKRP}
	\State $\Mz{L}{0:N-n-1} = {\Mz{X}{0:n-1}}^\Tra \cdot K_L$ 
	\State Partition $\Mz{L}{0}$ into $\rank$ column blocks of size $\SingleSize \times \RightSize$
	\For {$j \in \RangeSet{\rank}$}
		\State $\M{M}(:,j) \gets \Mzb{L}{0}{j} \cdot \M{K}_R(:,j)$ \Comment{$\Mzb{L}{0}{j}$ is column-major}
    \EndFor
	\Else \Comment{Use Right Partial MTTKRP}
        \State $\Mz{R}{0:n} = \Mz{X}{0:n} \cdot K_R$ 
        \State Partition $\Mz{R}{n}$ into $\rank$ column blocks of size $\SingleSize \times \LeftSize$
	\For {$j \in \RangeSet{\rank}$}
		\State $\M{M}(:,j) \gets \Mzb{R}{n}{j} \cdot \M{K}_L(:,j)$ \Comment{$\Mzb{R}{n}{j}$ is row-major}
    \EndFor
	\EndIf
  \EndProcedure
  \Ensure $\M[M]\ =\ \Mz{X}{n} \cdot (\Mn{U}{N-1} {\Khat} {\cdots} {\Khat} \Mn{U}{n+1} \Khat \Mn{U}{n-1} {\Khat} {\cdots} {\Khat} \Mn{U}{0})$ is $\SingleSize \times \rank$
  \end{algorithmic}
\end{algorithm}

\section{Experimental Results}
\label{sec:performance}


\subsection{Experimental Setup}

All experiments are benchmarked on a dual-socket Intel Xeon E5-2620 (Sandy Bridge) server with a total of 12 cores.
Each socket has a 15 MB L3 cache, and each core has a 256 KB L2 cache and 32 KB L1 data cache.
Each core has a clock rate of 2.00 GHz (with turbo boost disabled) and peak flop rate of 16 GFLOPS.
Our code is written in C and compiled with GCC version 5.4.0.
We use Intel's Math Kernel Library (MKL) version 2017.2.174.
The MATLAB benchmarks use version 9.0.0.341360 (R2016a) and Tensor Toolbox version 2.6.
All experiments are performed in double precision.

\subsection{KRP}

\begin{figure}
	\begin{subfigure}{\columnwidth}
		\includegraphics[width=\columnwidth]{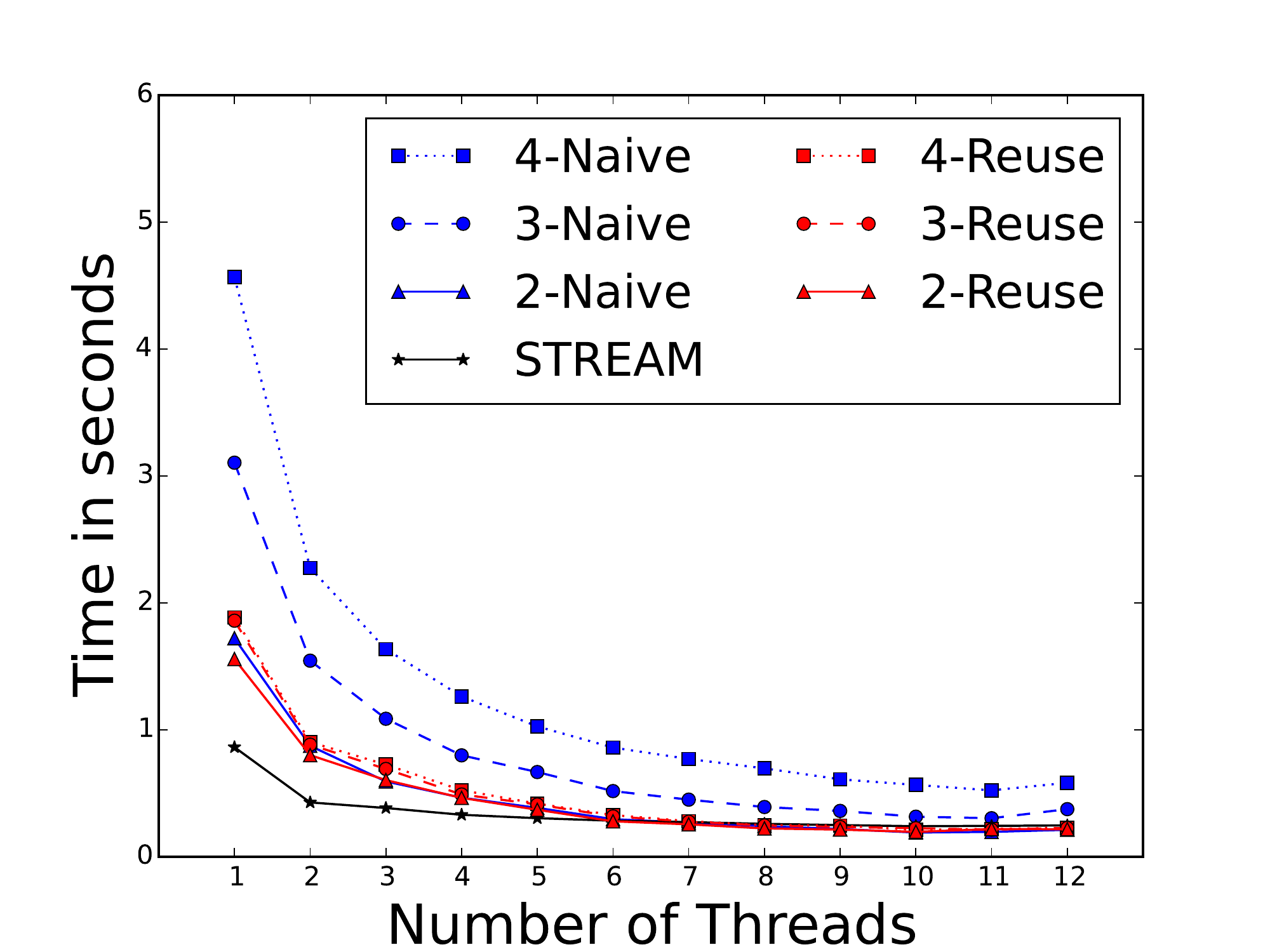}
		\caption{Output matrix with dimension $J \times 25$}
		\label{fig:KRPperf25}
	\end{subfigure} \\
	\begin{subfigure}{\columnwidth}
		\includegraphics[width=\columnwidth]{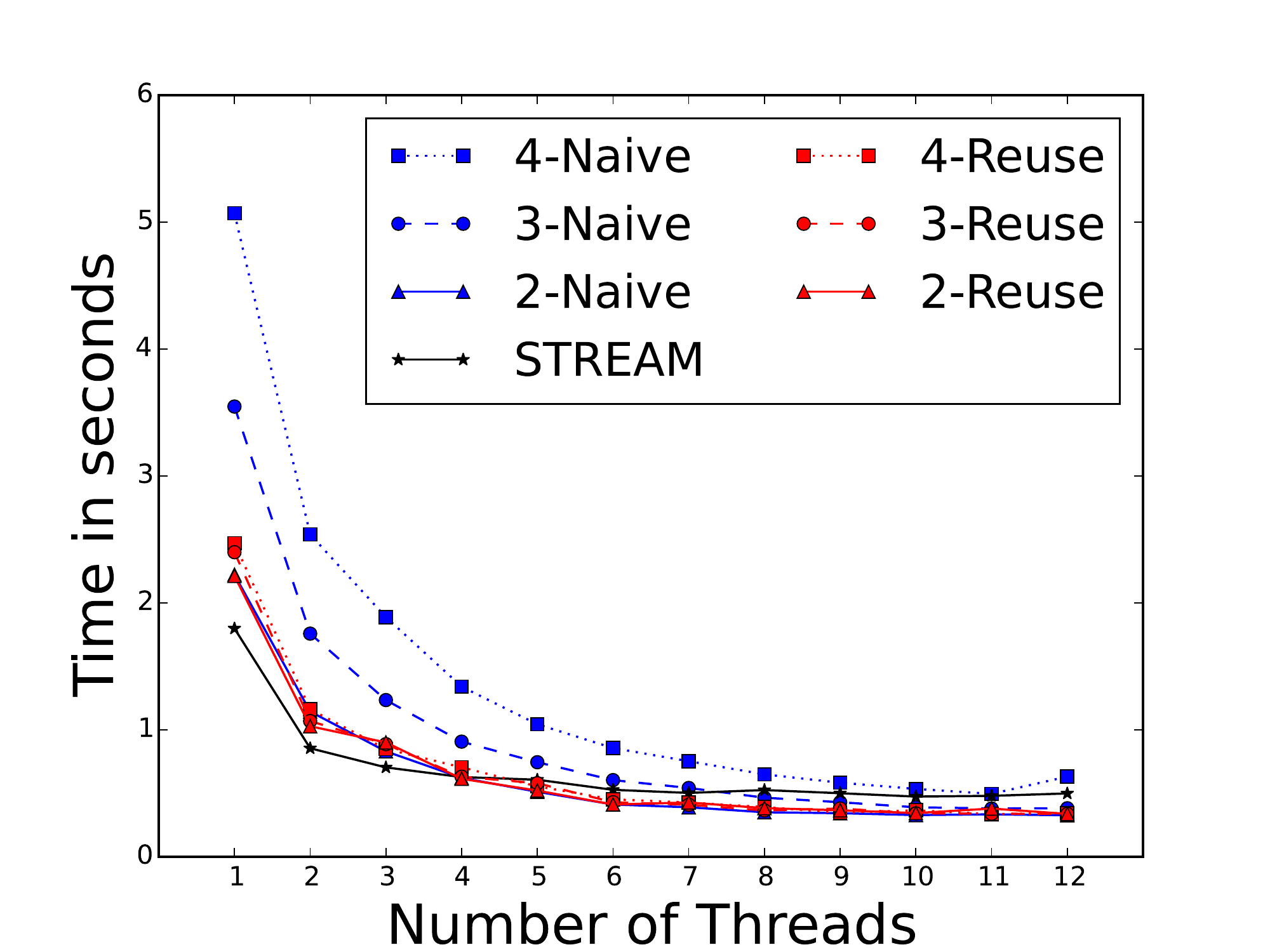}
		\caption{Output matrix with dimension $J \times 50$}
		\label{fig:KRPperf50}
	\end{subfigure}
\caption{Time comparison of \Cref{alg:KRP} against a naive algorithm and STREAM benchmark over varying numbers of threads.  Each experiment involves $J\approx 2e7$ output matrix rows and either 2, 3, or 4 input matrices and either 25 or 50 columns.  Both KRP algorithms compute a row of the output at time; \Cref{alg:KRP} avoids the redundant computation performed by the naive algorithm.}
\label{fig:KRPperf}
\end{figure}

We first consider the performance of \Cref{alg:KRP}, which computes the Khatri-Rao product of $Z$ input matrices.
\Cref{fig:KRPperf} presents performance results for \Cref{alg:KRP}, which exploits re-use of intermediate quantities, in comparison with a naive version of the algorithm and the STREAM benchmark \cite{McCalpin07}.
We consider $\rank=\{25,50\}$ (in \cref{fig:KRPperf25,fig:KRPperf50}) and in each case experiment with $Z=\{2,3,4\}$.
For each value of $\rank$ we choose the input matrix row dimensions to be all equivalent and such that their product is approximately 20 million.
This implies that for all experiments shown in \cref{fig:KRPperf25}, the output matrix has a size of roughly 500 million entries; in \cref{fig:KRPperf50} the output matrix has approximately 1 billion entries.

We measure the running time of three algorithms over various numbers of threads.
The results labeled ``Reuse'' correspond to \cref{alg:KRP}, which is the algorithm we use in the MTTKRP experiments.
The results labeled ``Naive'' correspond to a row-wise algorithm that does not store and re-use intermediate Hadamard products.
The STREAM benchmark we report is based on reading, scaling, and writing a matrix the same size as the output KRP matrix.
Each reported time is the average of 100 trials.

Our first conclusion from the data is that exploiting reuse is an important optimization for KRP.
\Cref{alg:KRP} outperforms its naive alternative, and the difference increases with $Z$ (note that for $Z=2$ there is no difference in algorithm).
For $Z=\{3,4\}$, the speedups of Reuse over Naive range from $1.5\times$ to $2.5\times$. \\

Our second conclusion is that \Cref{alg:KRP} is essentially a memory-bound operation, achieving competitive performance with the STREAM benchmark.
This is expected, as the number of flops in the optimized \Cref{alg:KRP} is the same as the number of output matrix entries.
However, because the input matrices are relatively small, we see that KRP can take even less time than STREAM (as in the case of $\rank=50$), which involves both a read and a write of the large matrix.

Finally, we see efficient scaling of our parallel variant of \Cref{alg:KRP}.
For $\rank=25$, we observe a parallel speedup range of $6.6-7.4\times$ for 12 threads; for $\rank=50$ the speedup range is $7.9-8.3\times$.

\subsection{MTTKRP}

In this section we discuss performance results for our proposed MTTKRP algorithms.
We compare the performance of 1-step (\Cref{alg:par-1s-MTTKRP}) and 2-step (\Cref{alg:2s-MTTKRP}) algorithms over various numbers of threads, noting that the two algorithms are equivalent for external modes ($n=0$ and $n=N{-}1$).
For a baseline, we also compare against the performance of a single BLAS call (MKL's implementation of DGEMM).
This benchmark is run on a single matrix multiplication between two column-major matrices that are the same size as the matricized tensor and the KRP, respectively.
It can be viewed as a lower bound on the performance of the most straightforward approach to MTTKRP (that reorders tensor entries) because it does not include the time required to reorder entries or form the explicit KRP.
Each reported result in this section is the median of 10 runs.

We note that in the case of the 1-step approach, the parallel algorithm (\cref{alg:par-1s-MTTKRP}) run with 1 thread is slightly different than the sequential algorithm (\cref{alg:seq-1s-MTTKRP}) for internal modes.
Instead of forming the full KRP $\M{K}$ explicitly, the parallel algorithm forms the left partial KRP and computes blocks of $\M{K}$ as needed.
Because we observed the parallel approach (when run with 1 thread) is slightly more efficient and uses less memory than the sequential approach, we use the parallel approach for all sequential benchmarks.

\subsubsection{Parallel Scaling}
\label{sec:parscaling}

\newcommand{\wsf}{.4\textwidth}
 \begin{figure*}
	\begin{subfigure}{\wsf}
		\centering
		\includegraphics[width=\textwidth]{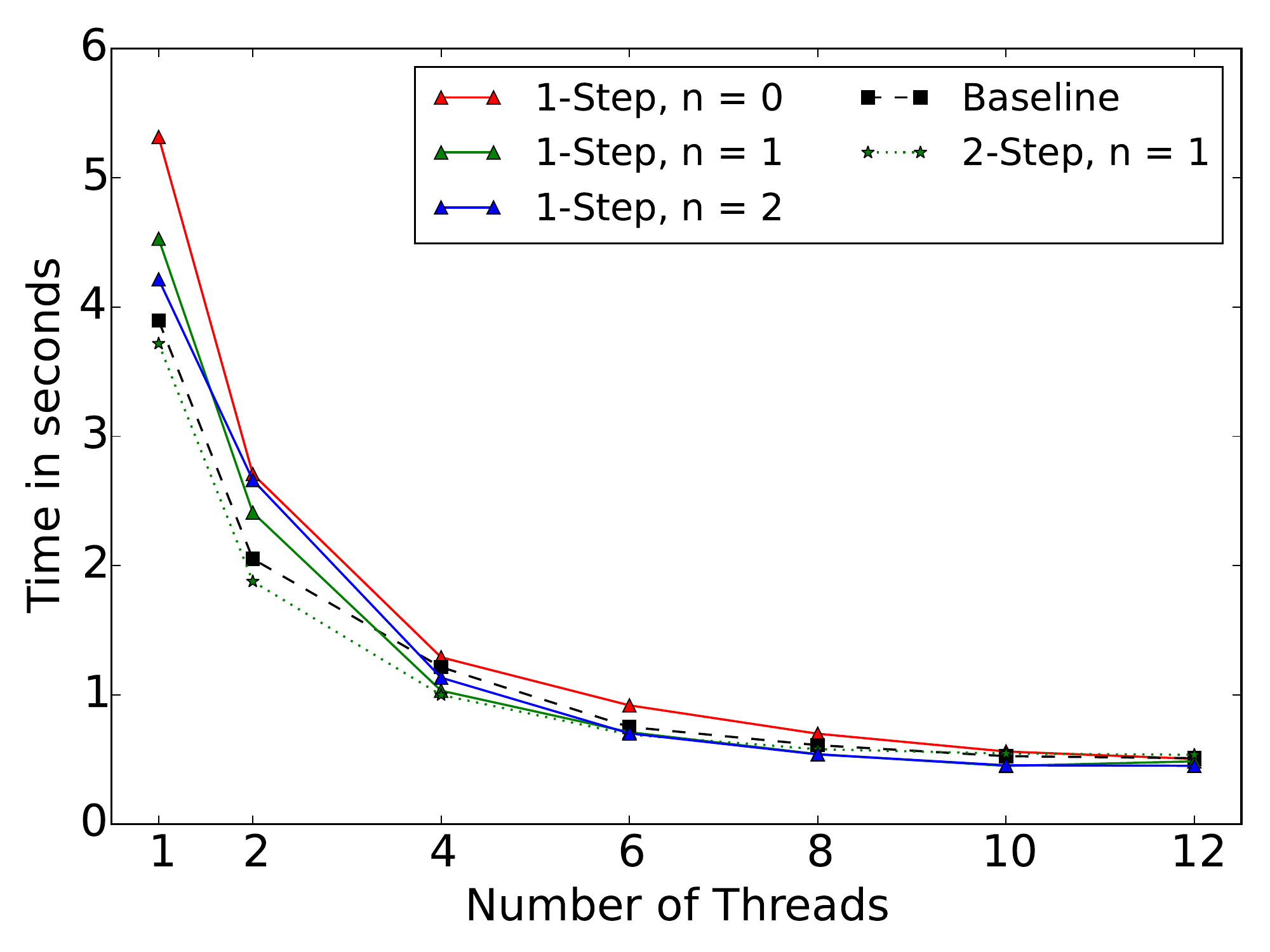}
		\caption{$N{=}3$: $900\times 900\times 900$}
	\end{subfigure} 
	\begin{subfigure}{\wsf}
		\centering
		\includegraphics[width=\textwidth]{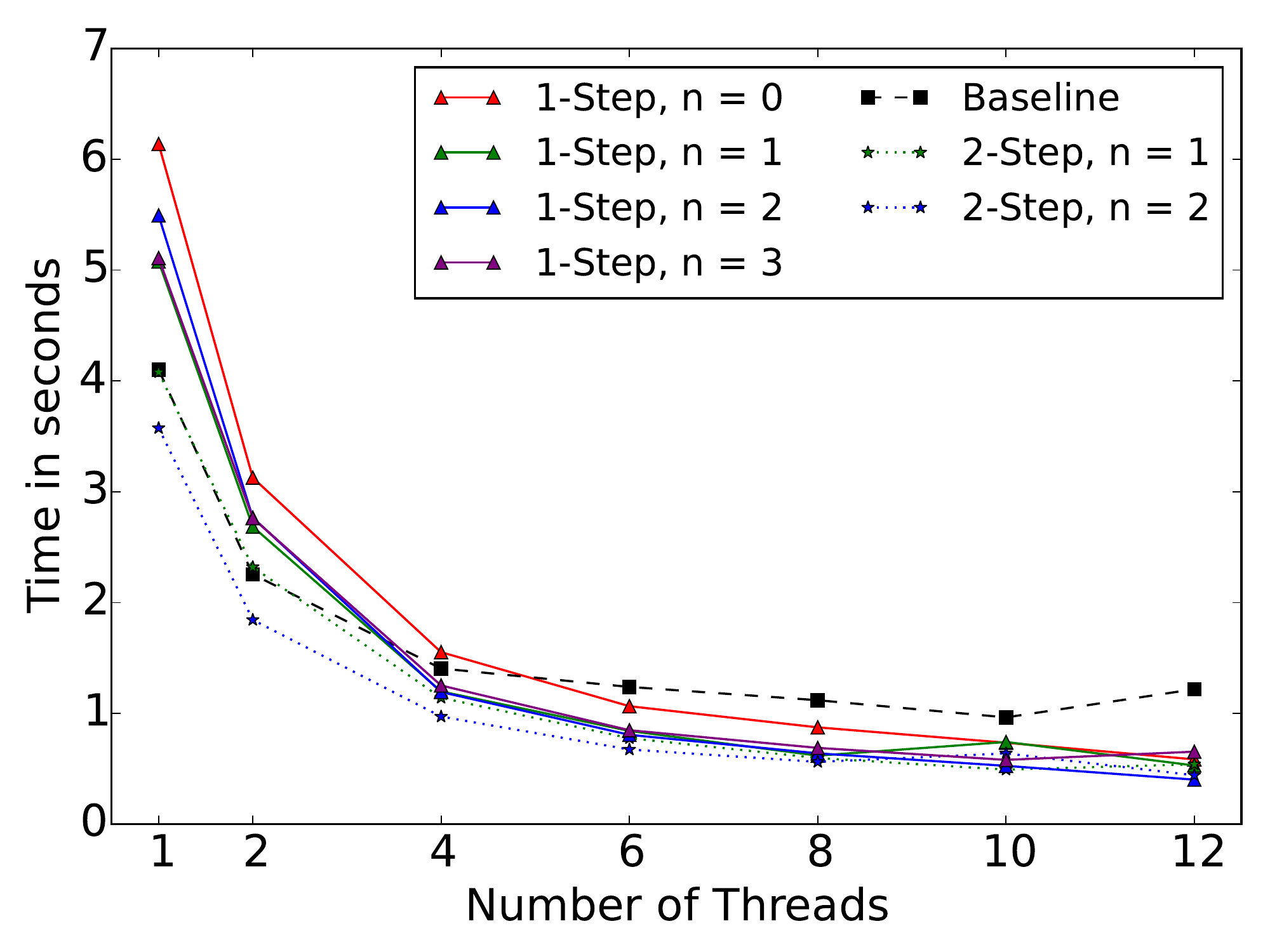}
		\caption{$N{=}4$: $165\times \cdots \times 165$}
	\end{subfigure}	\\
	\begin{subfigure}{\wsf}
		\centering
		\includegraphics[width=\textwidth]{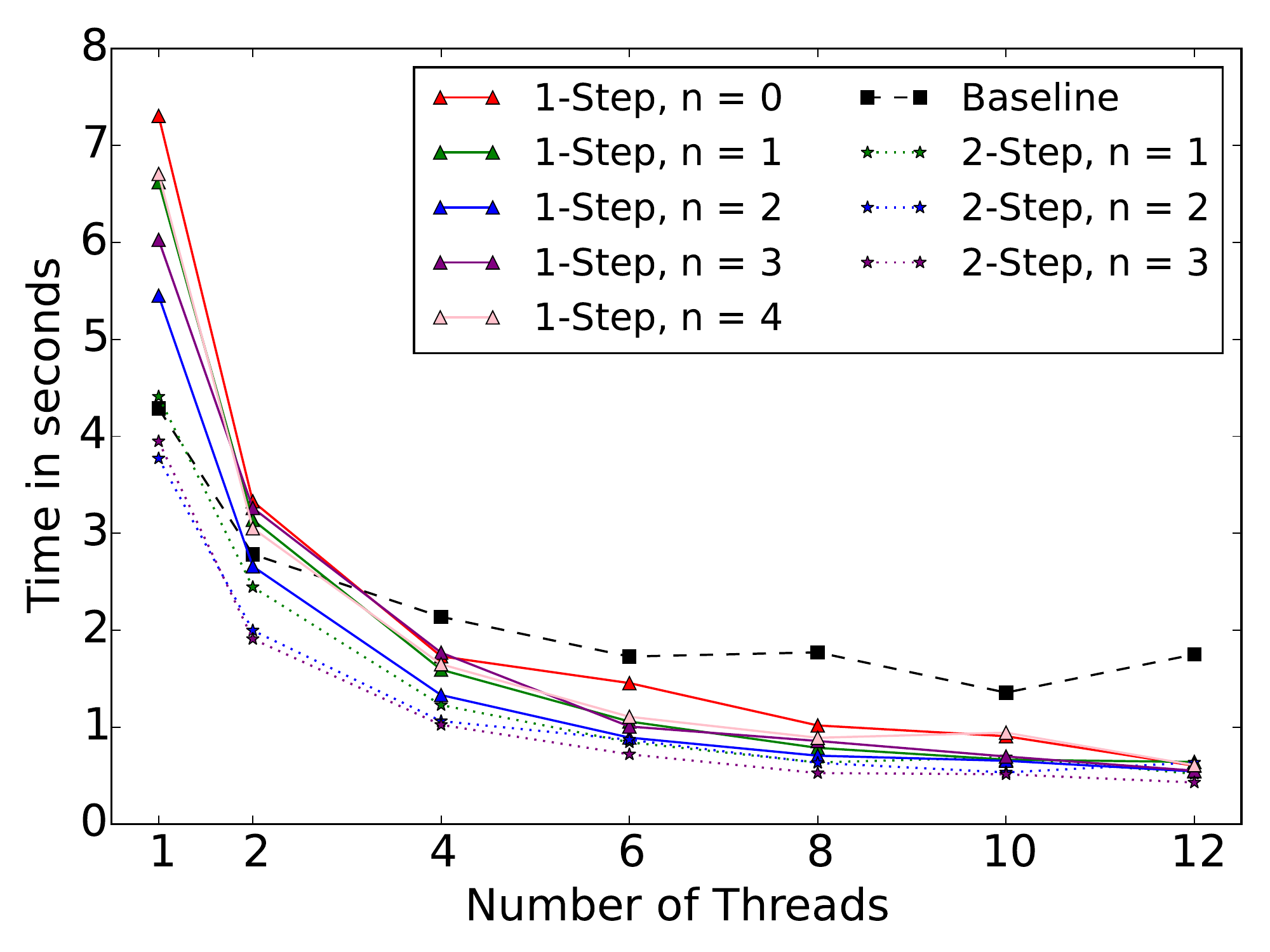}
		\caption{$N{=}5$: $60\times \cdots \times 60$}
	\end{subfigure}	
	\begin{subfigure}{\wsf}
		\centering
		\includegraphics[width=\textwidth]{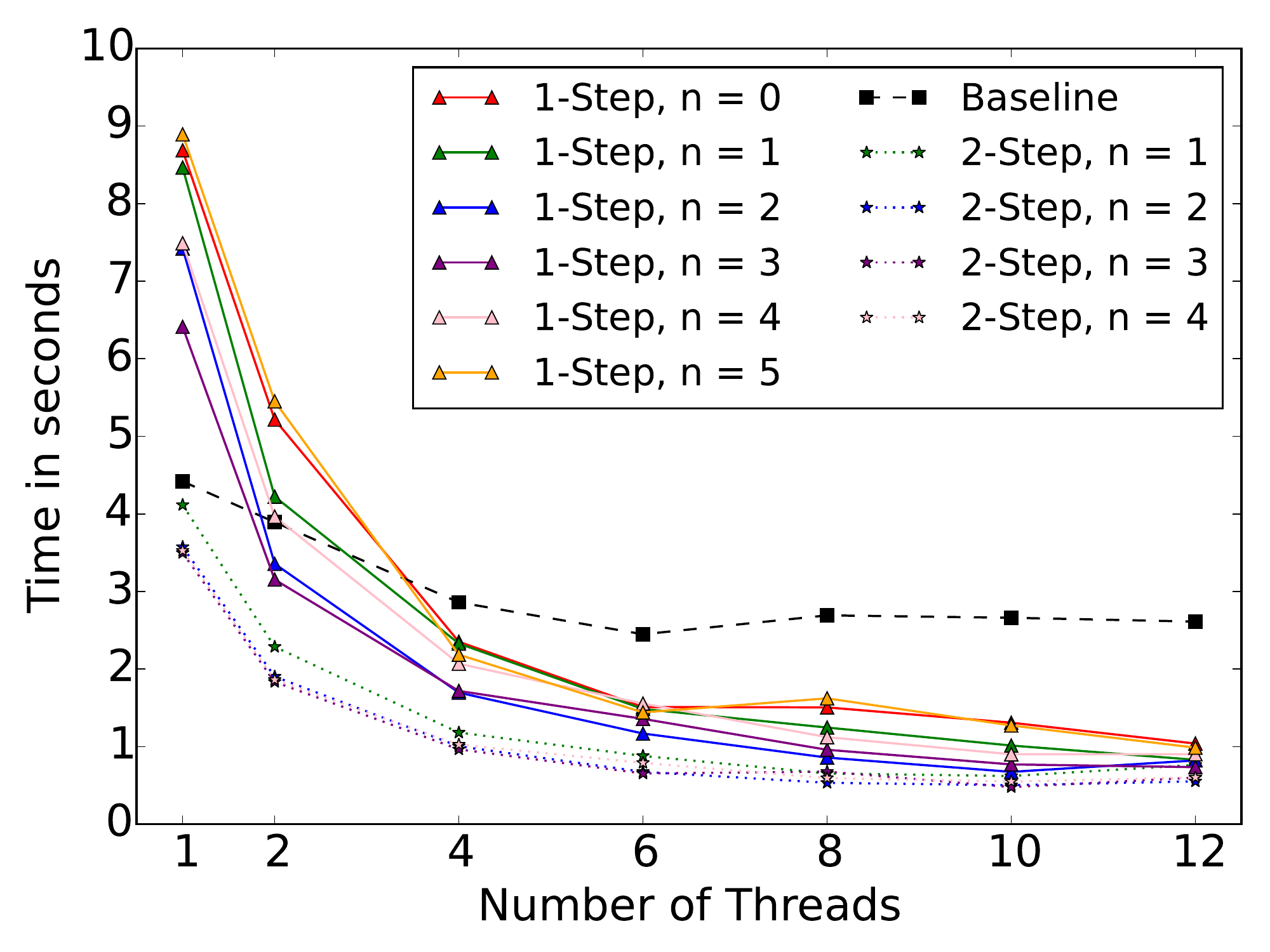}
		\caption{$N{=}6$: $30\times \cdots \times 30$}
	\end{subfigure}
	\caption{Time comparison of 1-step and 2-step MTTKRP algorithms for different modes over varying numbers of threads.
	Each subfigure corresponds to a number of modes; all experiments involve approximately 750 million tensor entries.
	The 2-step algorithm is defined only for inner modes.
	The baseline DGEMM benchmark is the time to multiply column-major matrices of the same dimensions as the MTTKRP. }
	\label{fig:parscaling}
\end{figure*}

We first consider the overall time and parallel scaling of each algorithm for different tensors.
\Cref{fig:parscaling} presents results for four different tensors, with $N=\{3,4,5,6\}$.
For each tensor, each dimension is the same and chosen so that the total number of tensor entries is approximately 750 million.
In each experiment, the number of columns in the output matrix is $\rank=25$.

Focusing first on sequential performance, \Cref{fig:parscaling} shows that the 2-step algorithm is faster than the baseline and that the 1-step algorithm is slower than the baseline, relationships that are consistent across both tensors and modes.
We note that the 2-step algorithm is not available for external modes.
In the worst case, the 1-step algorithm takes about $2\times$ as long as the baseline; the baseline is never slower than the 2-step algorithm by more than $25\%$ and never faster by more than $3\%$.
\Cref{sec:breakdown} further investigates the reasons for these relative performance differences.

Our next observation is that both 1-step and 2-step algorithms scale more efficiently than the baseline, particularly for larger $N$.
In fact, even at 4 threads, all of the proposed implementations are comparable or better than the single BLAS call, and they continue to improve up to 12 threads.
At 12 threads and for $N>3$, the speedup of 1-step and 2-step algorithms over the baseline range from $2\times$ to $4.7\times$, and the baseline still does not include time for reordering tensor entries or computing the KRP.

We believe part of the explanation for the poor scaling of the baseline implementation is that MKL has not fully optimized matrix multiplication of this shape, as has been observed in previous work \cite{DE+13}.
As $N$ increases in our benchmarks, the shape of the MTTKRP matrix multiplication approaches an inner product, with a long inner matrix multiplication dimension and a small output matrix.
The optimal parallelization of this computation involves write conflicts, for which we use temporary private memory and a parallel reduction, but MKL's implementation may be avoiding the memory footprint overhead of such an approach.

Unlike the baseline implementation, the 1-step and 2-step algorithms scale well to 12 threads.
The parallel speedup of the 1-step algorithm ranges from $8-12\times$ on 12 threads, and the 2-step parallel speedup ranges from $6-8\times$.
We note that the 2-step algorithm relies on the parallel performance of MKL, but it sees better parallel scaling than the baseline because the matricization within the partial MTTKRP is more square than that of the baseline approach (though it involves the same number of flops).
The fact that the 1-step algorithm scales slightly better than the 2-step algorithm implies that the parallel running times of the two approaches are fairly comparable at 12 threads.
We explore this further in \Cref{sec:breakdown}.

\subsubsection{Time Breakdown}
\label{sec:breakdown}

\renewcommand{\wsf}{.245\textwidth}
\begin{figure*}
	\includegraphics[width=.94\textwidth]{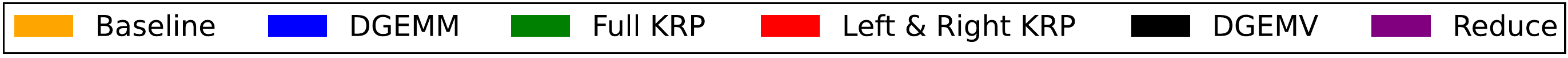}
	\begin{subfigure}{\wsf}
		\centering
		\includegraphics[width=\textwidth]{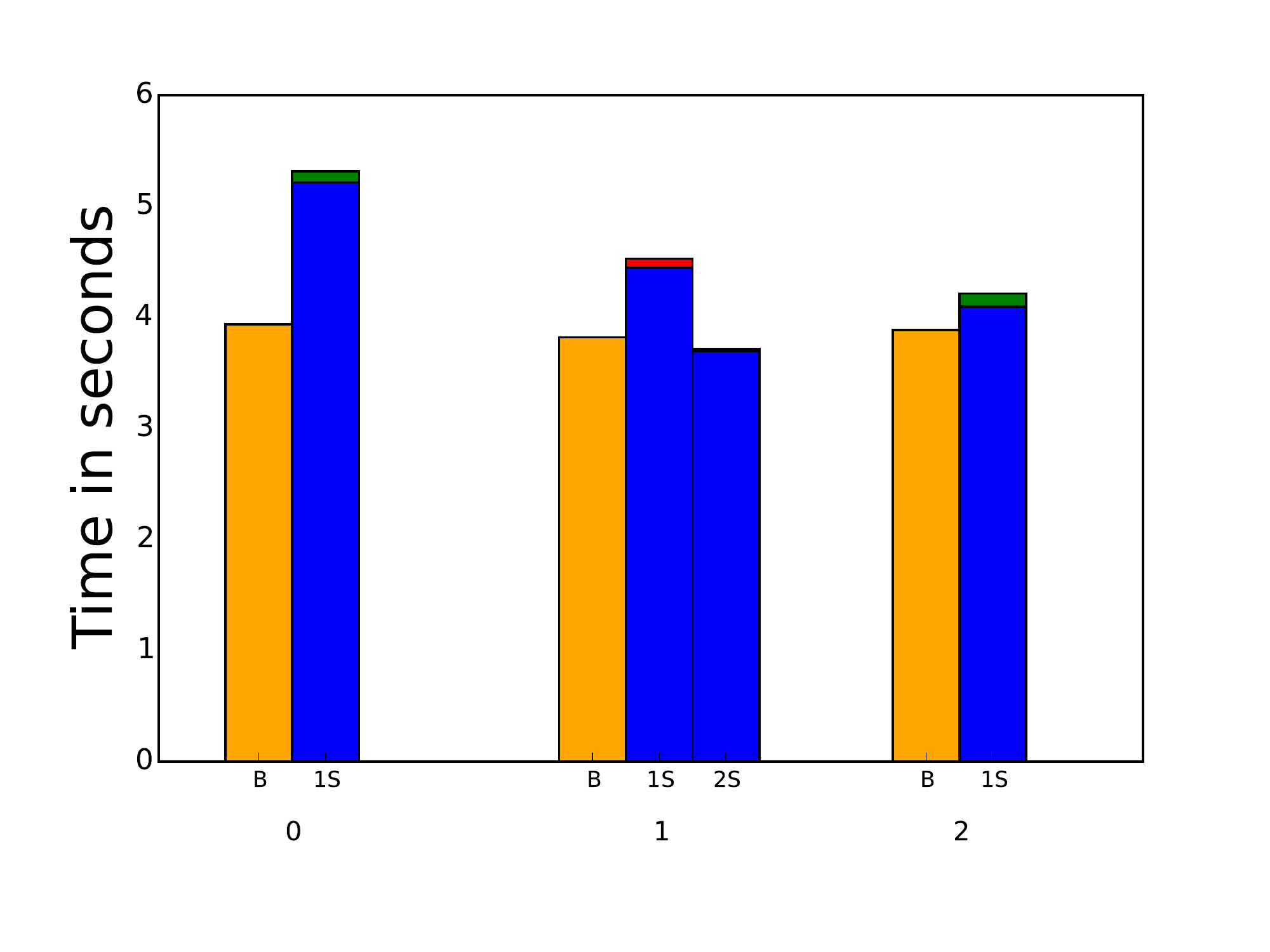}
		\caption{Seq. $N=3$}
	\end{subfigure}
	\begin{subfigure}{\wsf}
		\centering
		\includegraphics[width=\textwidth]{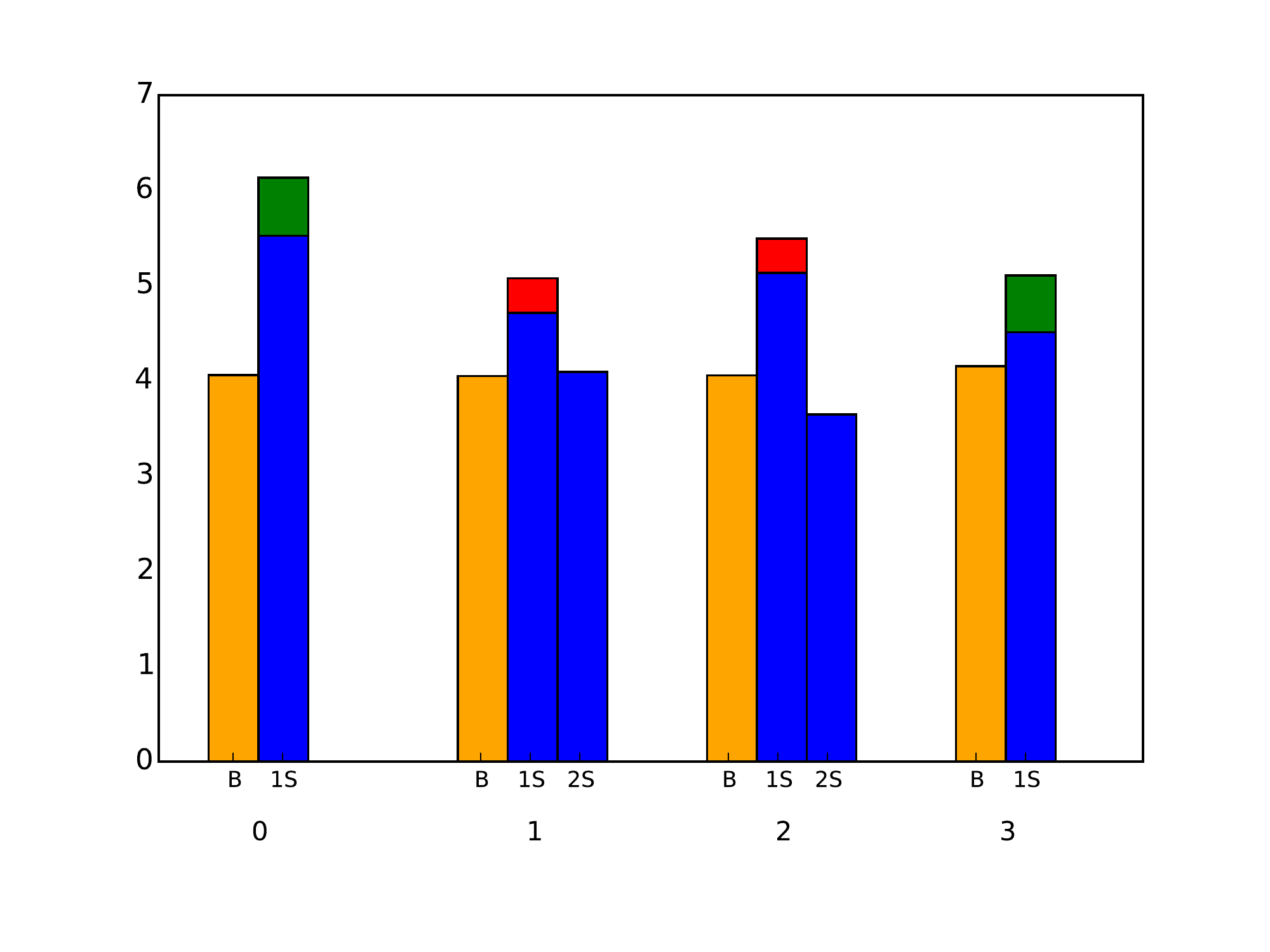}
		\caption{Seq. $N=4$}
	\end{subfigure}	
	\begin{subfigure}{\wsf}
		\centering
		\includegraphics[width=\textwidth]{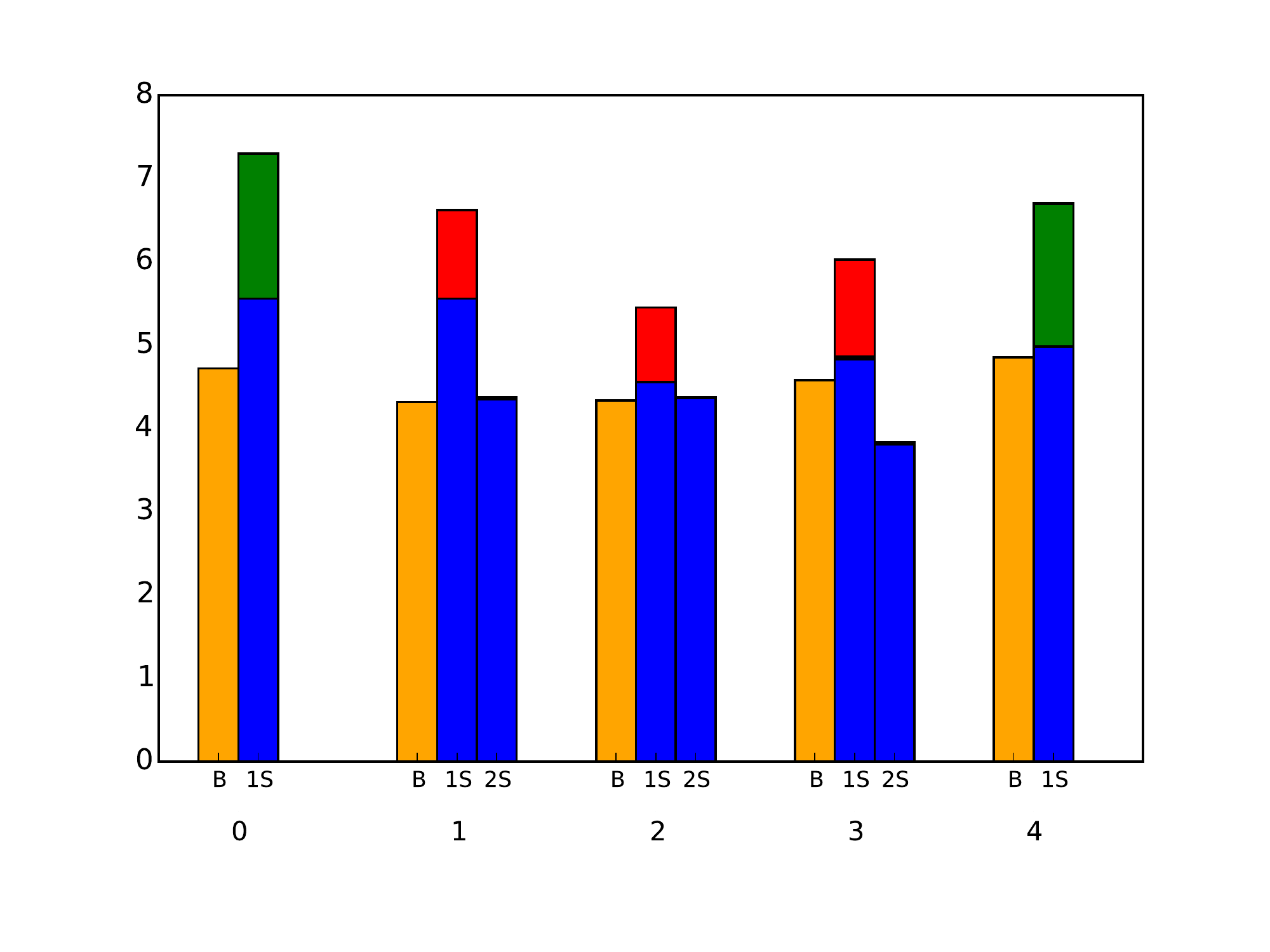}
		\caption{Seq. $N=5$}
	\end{subfigure}	
	\begin{subfigure}{\wsf}
		\centering
		\includegraphics[width=\textwidth]{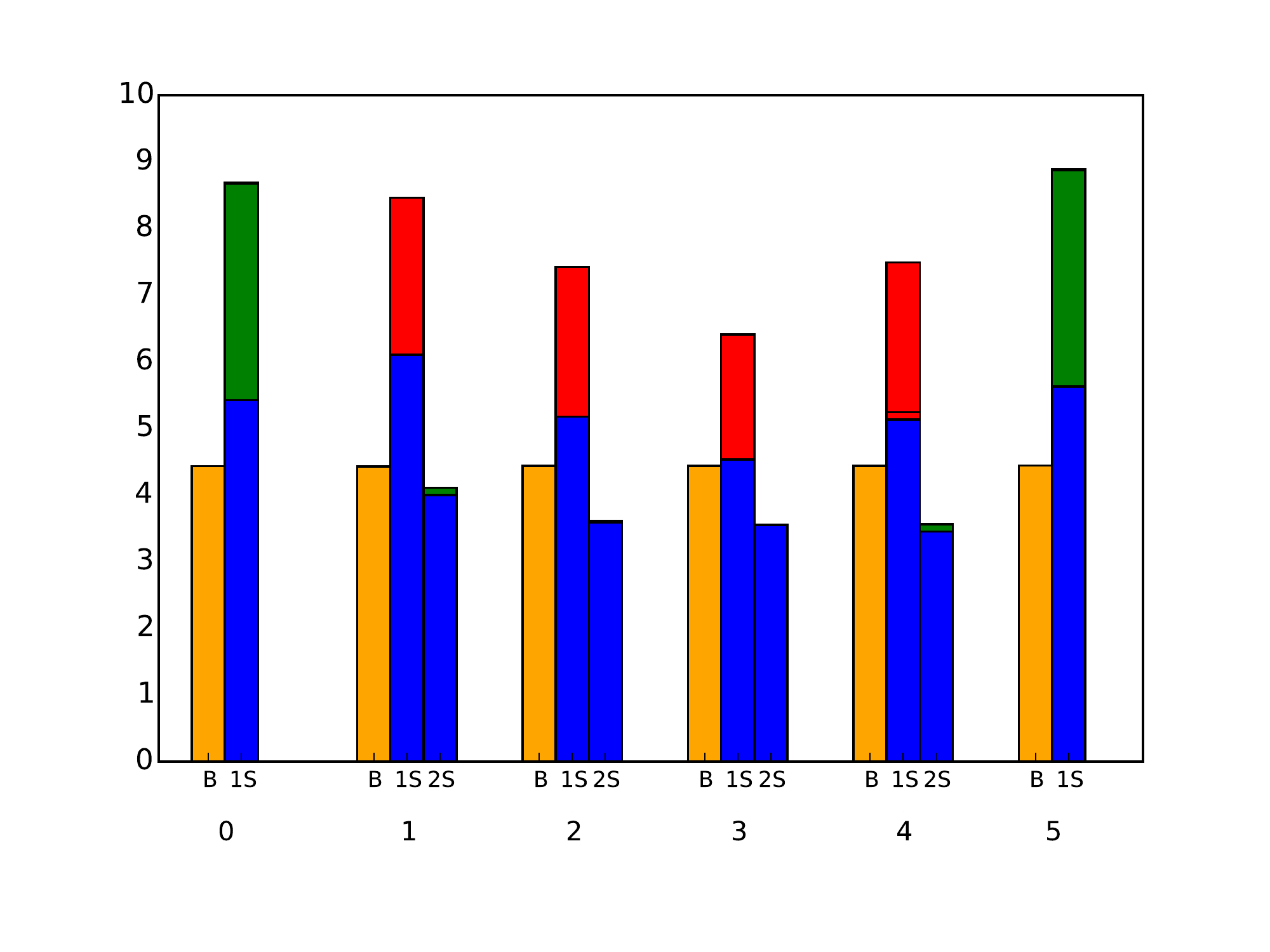}
		\caption{Seq. $N=6$}
		\label{fig:breakdown:S6}
	\end{subfigure}	\\
	\begin{subfigure}{\wsf}
		\centering
		\includegraphics[width=\textwidth]{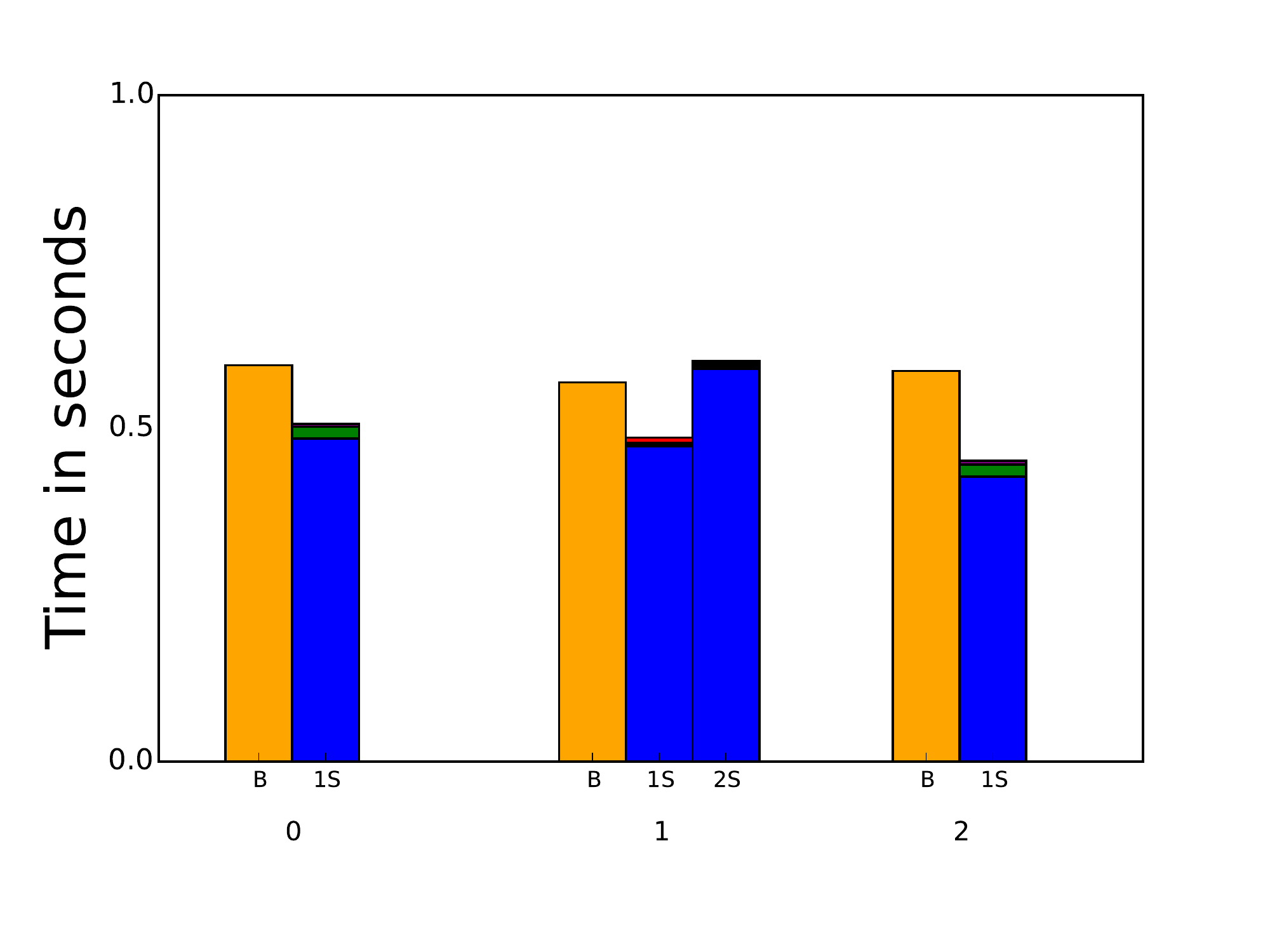}
		\caption{Par. $N=3$}
	\end{subfigure}
	\begin{subfigure}{\wsf}
		\centering
		\includegraphics[width=\textwidth]{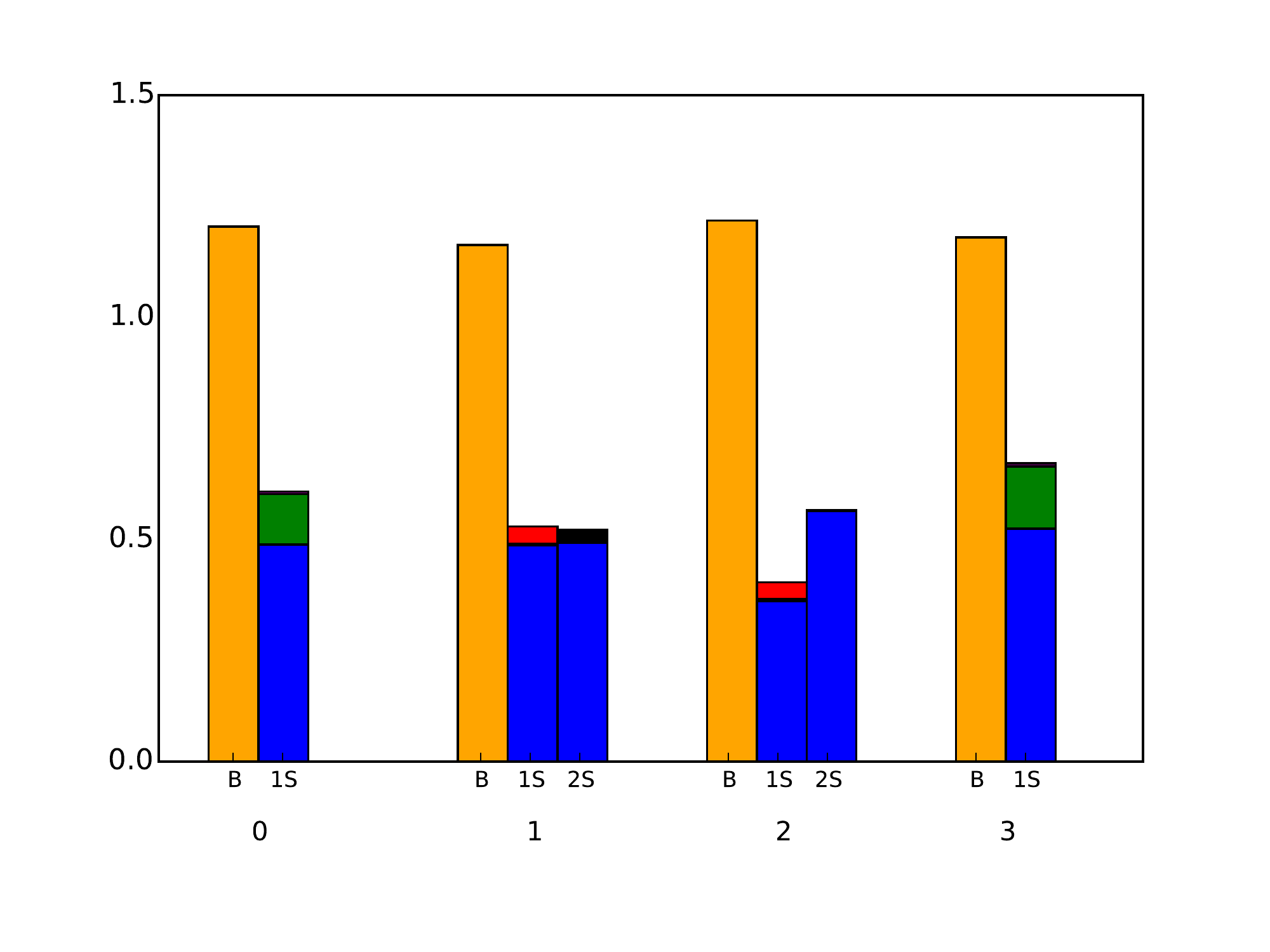}
		\caption{Par. $N=4$}
	\end{subfigure}	
	\begin{subfigure}{\wsf}
		\centering
		\includegraphics[width=\textwidth]{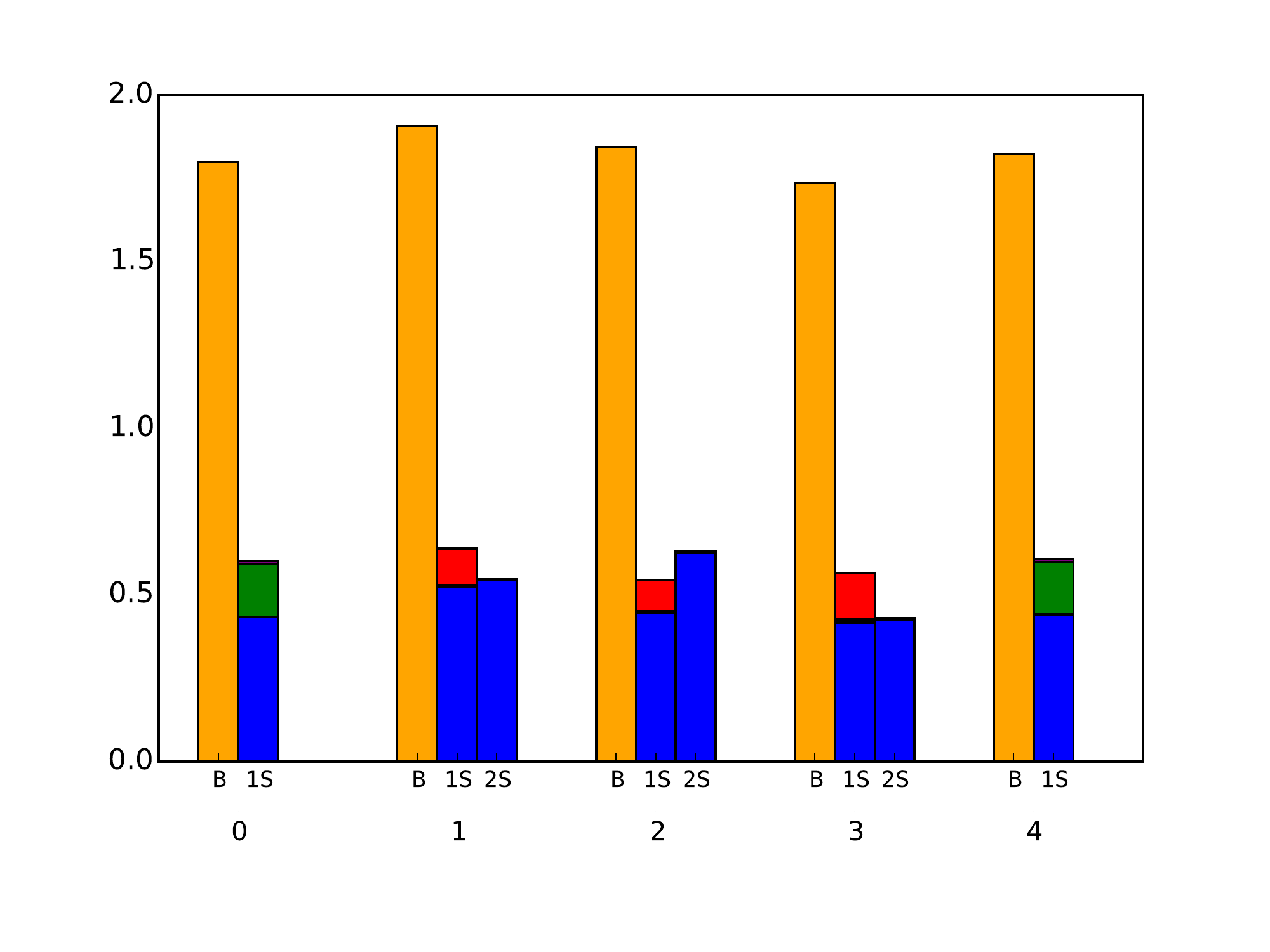}
		\caption{Par. $N=5$}
	\end{subfigure}	
	\begin{subfigure}{\wsf}
		\centering
		\includegraphics[width=\textwidth]{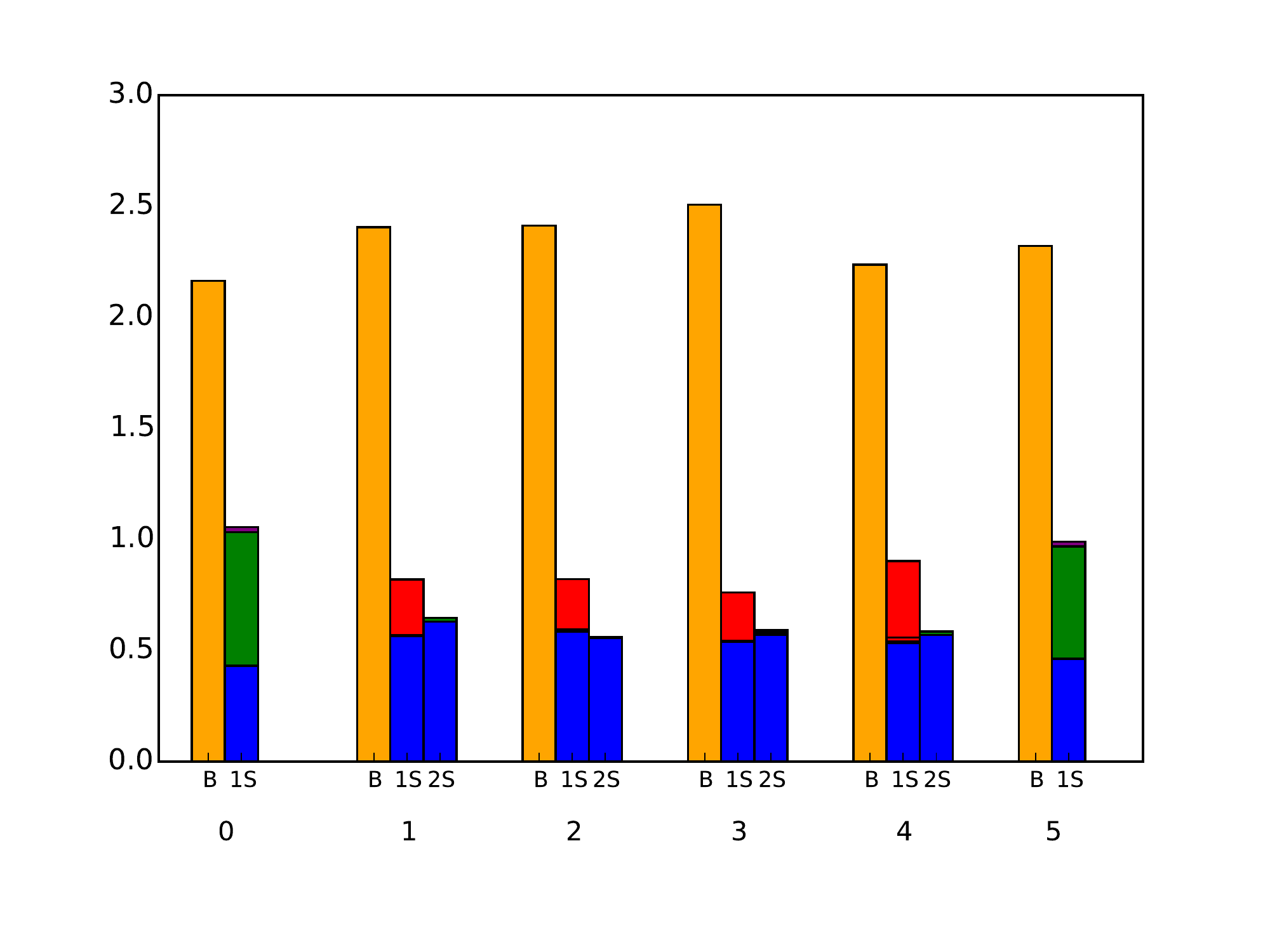}
		\caption{Par. $N=6$}
	\end{subfigure}	
	\caption{Time breakdown of 1-step and 2-step MTTKRP (and baseline DGEMM) across modes for varying numbers of modes.
	The top row corresponds to sequential time ($T=1$), and the bottom row corresponds to parallel time ($T=12$).
	Each column corresponds to a number of modes; all experiments involve approximately 750 million tensor entries and $\rank=25$ matrix columns.
	The baseline DGEMM benchmark is the time to multiply column-major matrices of the same dimensions as the MTTKRP.}
	\label{fig:breakdown}
\end{figure*}

\Cref{fig:breakdown} gives a detailed breakdown of the computation time of MTTKRP algorithms.
We consider the same four tensors as in \Cref{sec:parscaling}, with $N=\{3,4,5,6\}$, and we benchmark both sequential ($T=1$) and parallel ($T=12$) cases.
Each experiment uses $\rank=25$.

The baseline implementation has only one category: matrix multiplication (labeled Baseline in the legend).
The time for the 1-step algorithm (\cref{alg:par-1s-MTTKRP}) is broken down into matrix multiplication (DGEMM, line 8 or 16), forming the full KRP (Full KRP, line 7), forming the left KRP for internal modes and multiplying it by a row of the right KRP (Left \& Right KRP, lines 11 and 15), and performing the final parallel reduction (REDUCE, line 19).
The time for the 2-step algorithm (\cref{alg:2s-MTTKRP}) is broken down into matrix multiplication (DGEMM, line 5 or 11), matrix-vector multiplication (DGEMV, line 8 or 14), and forming the left and right KRPs (Left \& Right KRP, line 2 and 3).

Our first observation is that a considerable amount of time in the 1-step algorithm is spent computing the KRP, particularly for external modes.
For internal modes, only the left KRP is computed explicitly (which requires negligible time), and the rest of the KRP time is spent in computing blocks of the full KRP using a row of the right KRP.
In fact, this extra cost is the main reason the 1-step algorithm is slower than the baseline in the sequential case; the matrix multiplication time is generally comparable to the baseline even though it involves multiple BLAS calls for smaller matrix dimensions.
(Recall that the baseline ignores the cost of forming the KRP.)
Comparing the sequential 1-step performance to the parallel 1-step performance, we see that each category scales similarly and the proportions are generally preserved.

For internal modes, we observe that the 2-step algorithm spends almost all of its time in matrix multiplication.
The time spent in other computations is generally negligible.
Comparing the matrix multiplication time to the baseline, we see that the 2-step algorithm demonstrates slightly better performance because the matrix dimensions are more amenable for MKL.
In the sequential case, the 2-step algorithm is generally faster than the 1-step algorithm (for internal modes).
In the parallel case, the two algorithms are comparable, though the 2-step algorithm is usually slightly faster.

\subsubsection{Neuroscience Data}

The underlying motivation for this work is to speed up CP-ALS in order to analyze neuroscience (fMRI) data.
Our data is a 4-way tensor of size $225 \times 59 \times 200 \times 200$, representing for 225 time steps and for 59 subjects the correlation between fMRI signals measured at 200 different brain regions.
For a different type of analysis, we linearize the last two modes; because the tensor is symmetric in these two modes this linearization also reduces the number of tensor entries by a factor of 2. 
The corresponding 3-way tensor is $225 \times 59 \times 19900$.

In this section, we compare the performance of Matlab code using the Tensor Toolbox with our implementation of CP-ALS using efficient MTTKRP kernels.
For dense tensors, the current available software packages (e.g., N-way Toolbox \cite{AB00}, Tensor Toolbox \cite{TensorToolbox}, Tensorlab \cite{Tensorlab}) are all written in Matlab.
Because the Matlab software packages cast computations as matrix operations, and Matlab uses efficient BLAS implementations like MKL, we can expect reasonable performance from Matlab.
However, on multicore servers, the only opportunity for parallelization in the packages is within BLAS calls.

\begin{figure}
	\includegraphics[width=\columnwidth]{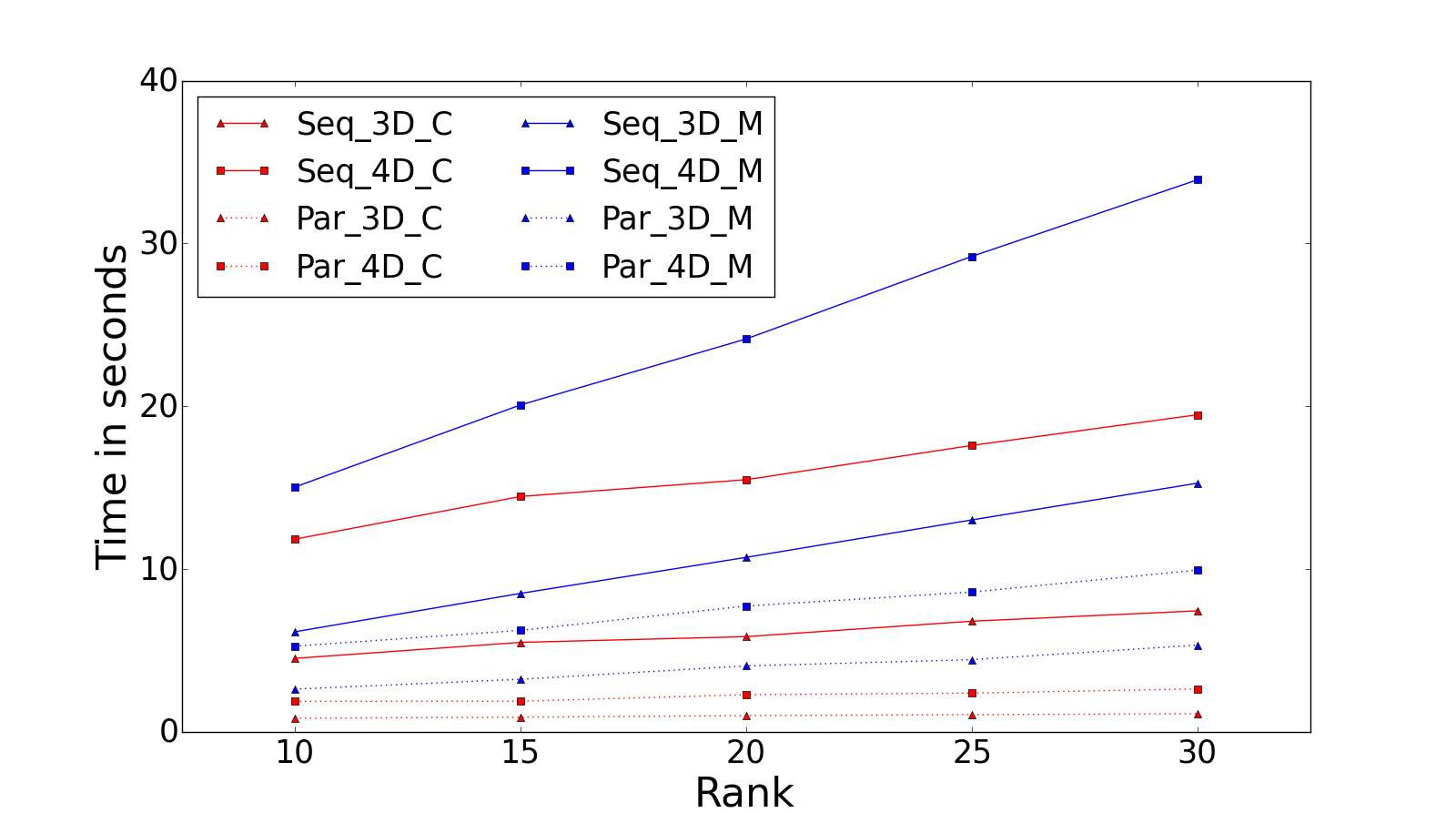}
	\caption{Per-iteration CP-ALS time for Matlab and C implementations over different ranks. The Matlab implementation is Tensor Toolbox's \texttt{cp\_als} function, and the C implementation is ours, using the most efficient 1-step and 2-step MTTKRP algorithm for each mode. Parallel runs are given all 12 cores on the machine.  Tensor sizes are $225\times59\times200\times200$ (4D) and $225\times59\times19900$ (3D).}
	\label{fig:CPALSperf}
\end{figure}

\Cref{fig:CPALSperf} shows the per-iteration run times for computing CP decompositions on our 3D and 4D application tensors with $\rank = \{10,15,20,25,30\}$, using both sequential and parallel, MATLAB and C implementations.
Our C implementation of CP-ALS employs \Cref{alg:par-1s-MTTKRP} (1-step) for both outer modes and \Cref{alg:2s-MTTKRP} (2-step) for all inner modes.
We observe up to a $2\times$ speedup of our sequential implementation over Matlab, running on only 1 core.
In the parallel case, the highest speedup of our implementation over Matlab (running with all 12 cores available) comes for the largest $\rank$: $6.7\times$ for the 3D tensor and $7.4\times$ for the 4D tensor.

\Cref{fig:breakdown-fMRI} gives the time breakdown for our implementation of MTTKRP for the application tensors, which have varying dimensions across modes.
This plot can be contrasted with \Cref{fig:breakdown}, which depicts data for synthetic tensors that have all the same dimensions across modes.
In particular, note that the KRP cost is relatively more significant in small modes ($n=1$, $I_1=59$), which agrees with the larger ratio of flops.
We observe that for large modes, the 1-step algorithm is competitive with the baseline in the sequential case and again outperforms the baseline in the parallel case.
The 2-step algorithm is consistently better than the baseline, and significantly better in the parallel case.
For mode $n=1$ the parallel MTTKRP algorithms are $2.8\times$ and $3.5\times$ faster than the baseline for 3D and 4D, respectively.

\renewcommand{\wsf}{.49\columnwidth}
 \begin{figure}
	\includegraphics[width=.94\columnwidth]{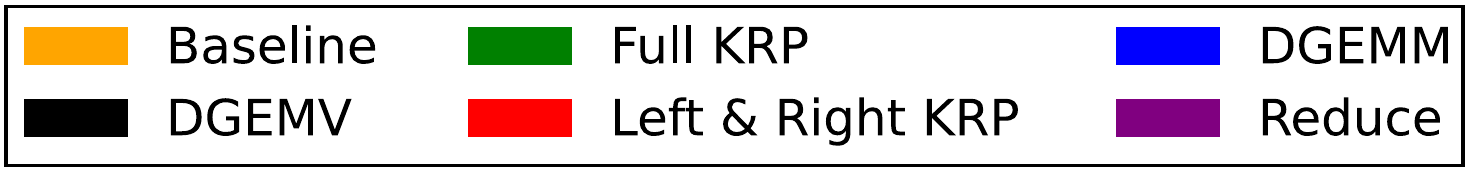}
	\begin{subfigure}{\wsf}
		\centering
		\includegraphics[width=\textwidth]{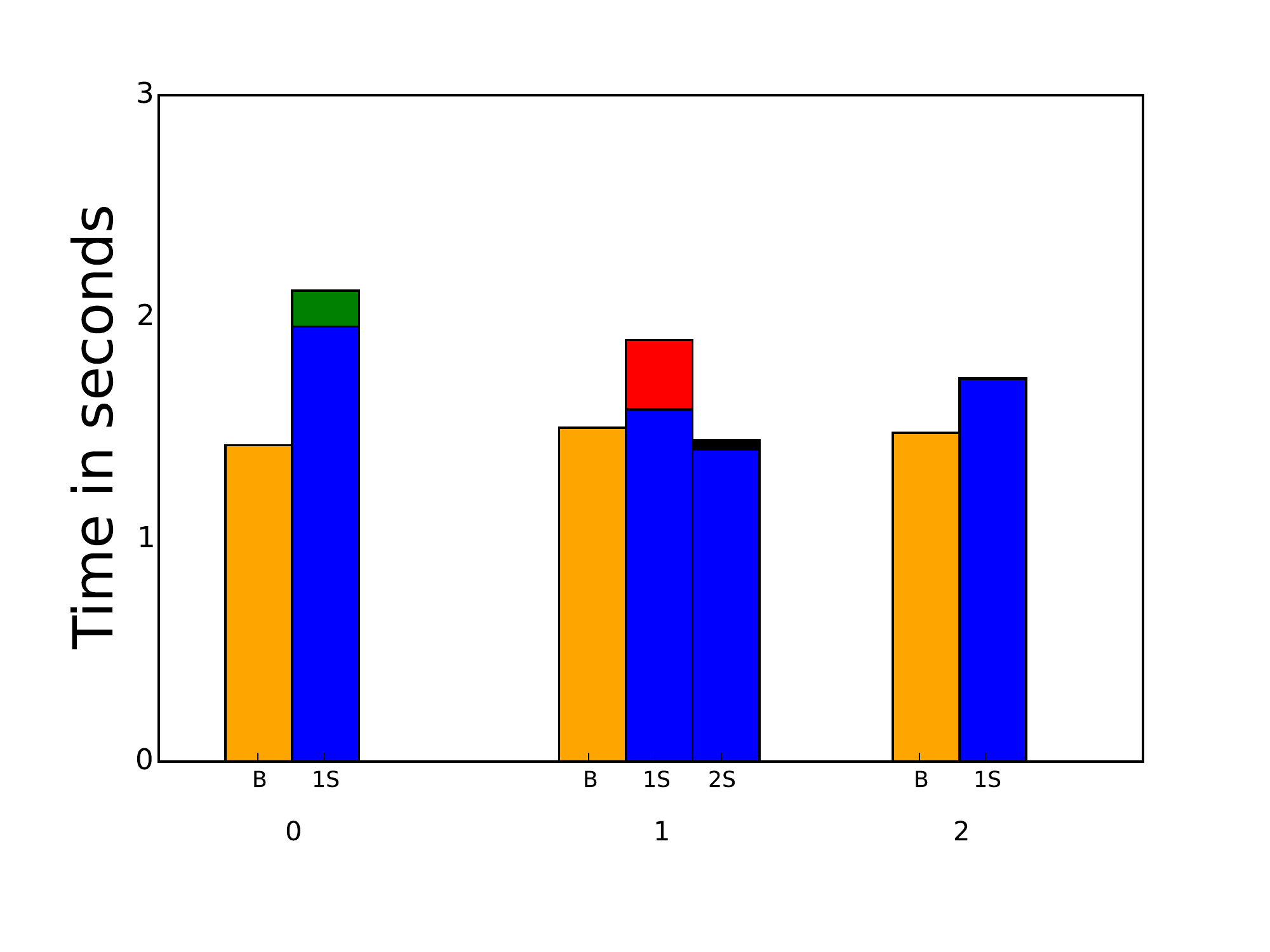}
		\caption{Seq. $N=3$}
		\label{fig:breakdown-fMRI:S3}
	\end{subfigure}
	\begin{subfigure}{\wsf}
		\centering
		\includegraphics[width=\textwidth]{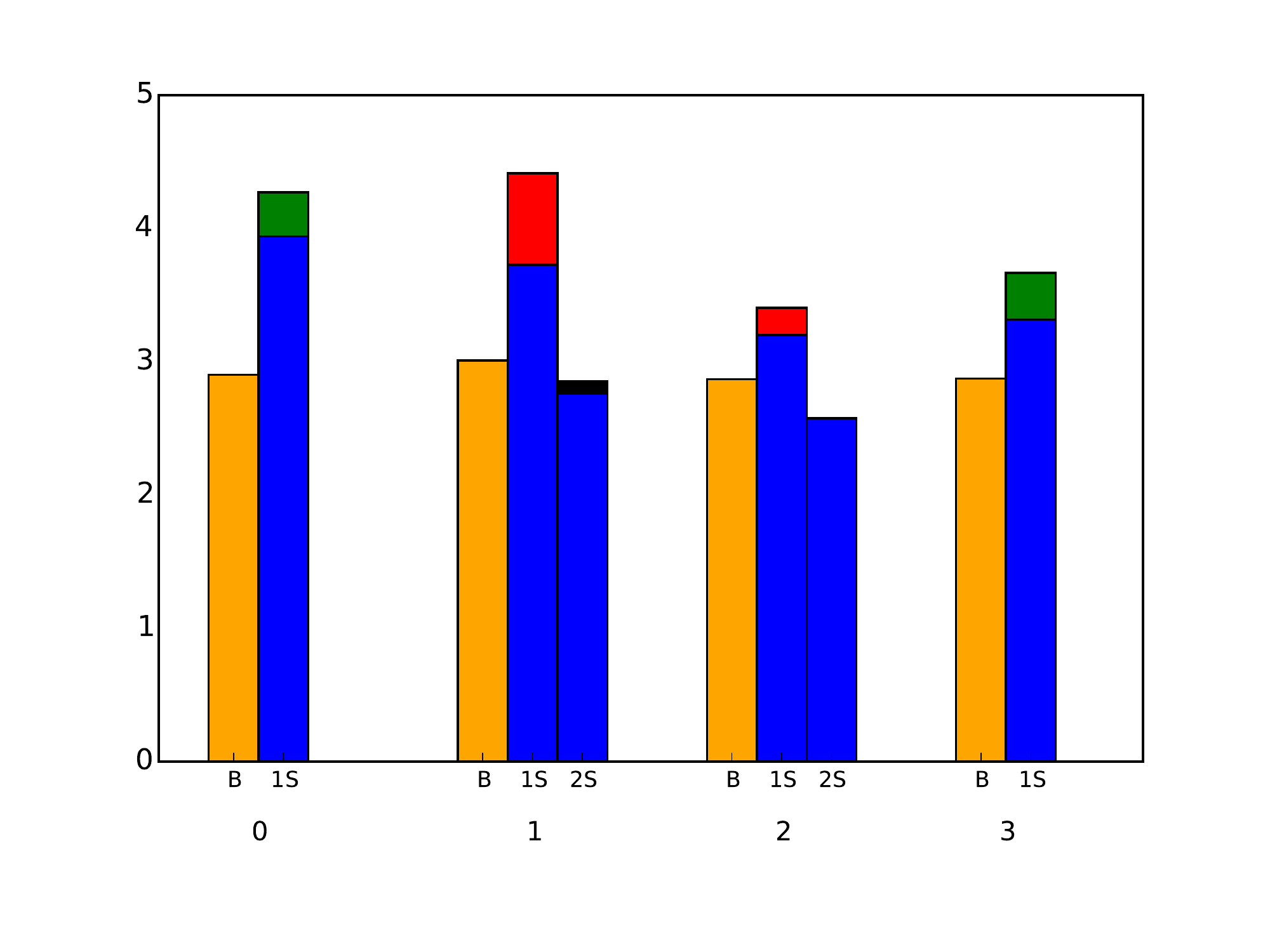}
		\caption{Seq. $N=4$}
		\label{fig:breakdown-fMRI:S4}
	\end{subfigure} \\
	\begin{subfigure}{\wsf}
		\centering
		\includegraphics[width=\textwidth]{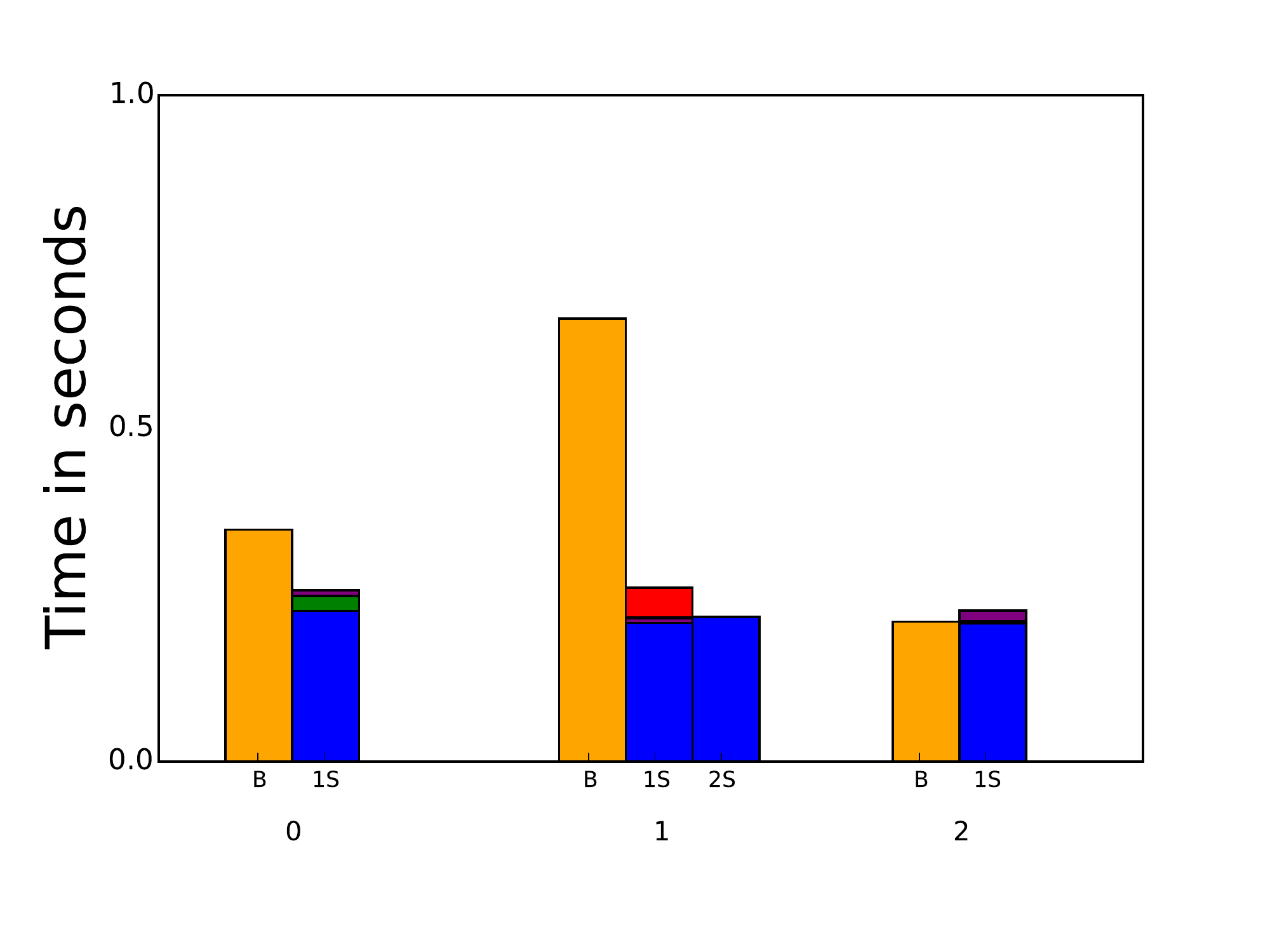}
		\caption{Par. $N=3$}
		\label{fig:breakdown-fMRI:P3}
	\end{subfigure}
	\begin{subfigure}{\wsf}
		\centering
		\includegraphics[width=\textwidth]{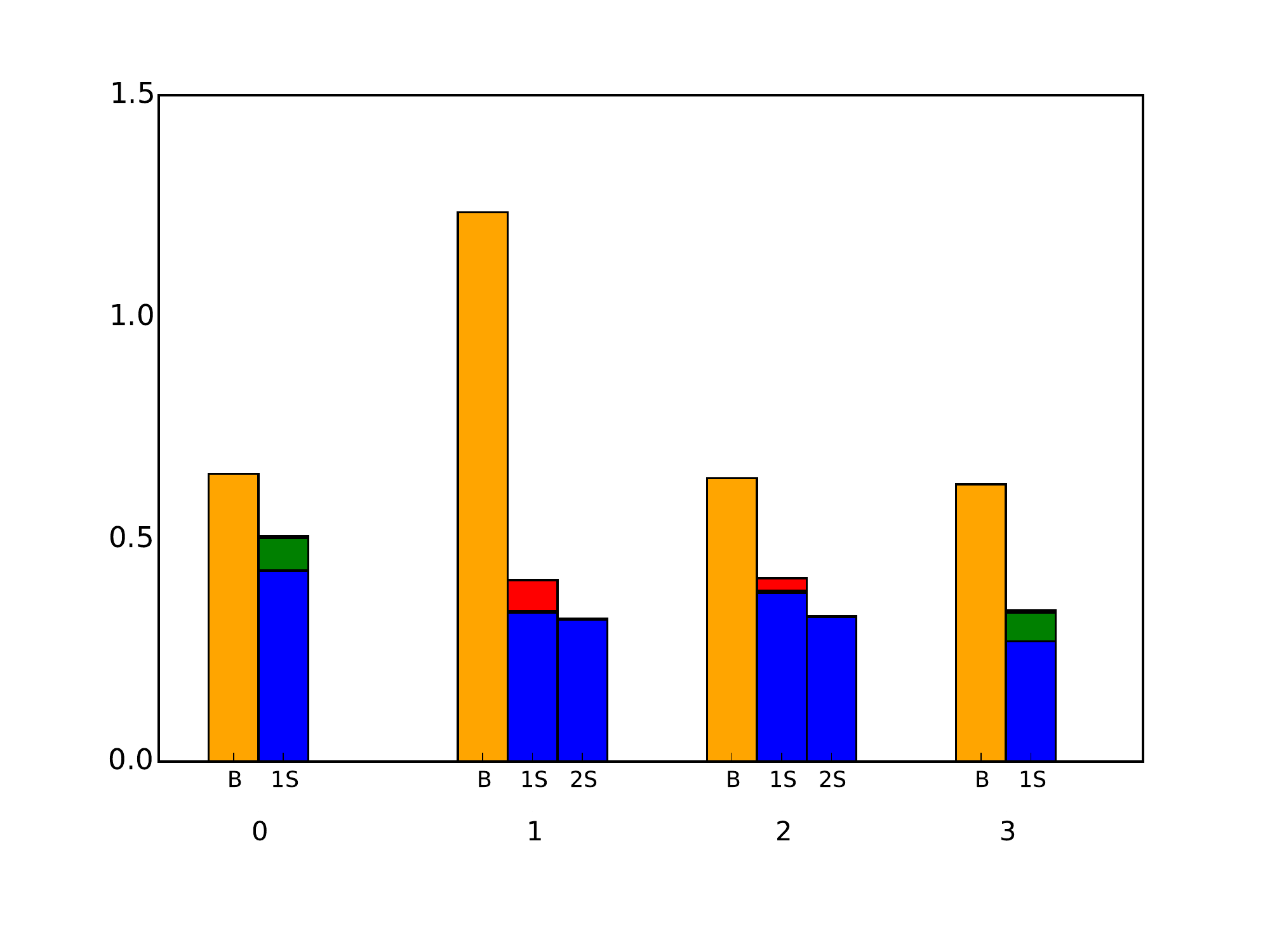}
		\caption{Par. $N=4$}
		\label{fig:breakdown-fMRI:P4}
	\end{subfigure}	
	\caption{Time breakdown of 1-step and 2-step MTTKRP (and baseline DGEMM) across modes for 3D and 4D fMRI tensors.
	The top row corresponds to sequential time ($T=1$), while the bottom row corresponds to parallel time ($T=12$).
	The baseline DGEMM benchmark is the time to multiply column-major matrices of the same dimensions as the MTTKRP.}
	\label{fig:breakdown-fMRI}
\end{figure}

\section{Conclusion}
\label{sec:conclusion}

In this paper, we introduce a parallel algorithm for KRP and two parallel algorithms for MTTKRP. 
Our performance results indicate that our implementations perform well sequentially and scale efficiently up to 12 threads.
We also show that improving the performance of these kernels results in faster CP-ALS iterations for application problems.

One conclusion that we wish to highlight from the performance results is the high relative cost of the KRP computation in the 1-step algorithm.
For example, in the case of the external modes of synthetic 6-way tensor where each mode has dimension 30 (see \cref{fig:breakdown:S6}), the KRP takes about a third to a half of the time even though the number of flops is $1/30$th the number of flops involved in the matrix multiplication.
This is due in large part to the memory boundedness of the KRP computation.
Just as tensor reordering should be avoided, future optimization of MTTKRP should avoid computing large KRPs.

The natural next step for this work is to implement the algorithm proposed by Phan \emph{et al.} \cite[Section III.C]{PTC13} for avoiding recomputation across MTTKRPs of different modes, which can be done for CP-ALS and other gradient-based optimization methods.
In particular, the computational kernels of the full algorithm are the same as the single-mode computation.
Using this algorithm, we could expect a further reduction in per-iteration CP-ALS time of around $50\%$ in the 3D case and $2\times$ in the 4D case (and higher for larger $N$).
We note that this algorithm also benefits from avoiding computing large KRPs.

We have also noticed that with improvements of MTTKRP performance, other computations within CP-ALS, such as the residual error computation, have become relatively more costly.
Improving the efficiency of these new bottlenecks could yield overall performance increases. 

\newpage

\bibliographystyle{ACM-Reference-Format}
\bibliography{paper}

\end{document}